\documentclass[12pt,a4paper]{article}

\usepackage{amsmath, amssymb, amsthm}
\usepackage{graphicx}
\usepackage{tikz}
\usepackage{setspace}
\usepackage{geometry}
\usepackage{titlesec}
\usepackage{hyperref}
\usepackage{footnote}
\usepackage{endnotes}
\usepackage{natbib}
\usepackage{lmodern} 
\usepackage[T1]{fontenc}
\usepackage{caption}
\usepackage{float}
\usepackage{bm}

\usepackage[utf8]{inputenc}
\usepackage{graphicx}
\usepackage{amsmath}
\usepackage{setspace}
\usepackage{geometry}
\usepackage{times}
\usepackage{hyperref}
\usepackage{natbib}
\usepackage{amsfonts}
\usepackage{authblk} 
\titleformat{\chapter}[hang]{\LARGE\bfseries}{\thechapter\quad}{0pt}{}

\begin{document}














\title{\textbf{Large Wave Direction Data Modeling Using Wrapped Spatial Gaussian Markov Random Fields}}
\author{Arnab Hazra}
\affil{Department of Mathematics and Statistics, Indian Institute of Technology Kanpur, Kanpur 208016, India}
\date{}



\maketitle

\begin{abstract}
\noindent 
Statistical modeling of dependent directional data remains relatively underexplored, particularly in high-dimensional spatial settings. Existing approaches for spatial angular data primarily rely on wrapped Gaussian process (WGP) models, which provide a coherent framework for capturing spatial dependence on the circle. However, WGP-based methods become computationally challenging when the spatial domain is large, and observations are available at high resolution. This limitation is especially relevant in the analysis of large-scale geological and climate phenomena, such as tsunamis and hurricanes, where directional measurements (e.g., wave or wind directions) may be available over an entire ocean basin. To address these challenges, we propose a wrapped Gaussian Markov random field (WGMRF) model for large spatial directional datasets. By exploiting the sparse precision structure inherent in Gaussian Markov random fields, the proposed approach achieves substantial computational gains while preserving flexible spatial dependence on the circular scale. We discuss key properties of the model, including its identifiability and dependence characteristics. The model fitting involves standard Markov chain Monte Carlo techniques. Through extensive simulation studies and an application to the wave direction data across the Indian Ocean during the 2004 Indian Ocean Tsunami, we compare the proposed method with both a non-spatial wrapped Gaussian model and a low-rank WGP alternative. The results demonstrate that the WGMRF offers improved predictive performance and scalability in large-domain applications.
\end{abstract}


\section{Introduction}
Wave direction is a fundamental component of ocean dynamics, governing how wave energy propagates across basins and interacts with coastlines. The direction of incoming waves strongly influences coastal erosion, sediment transport, shoreline evolution, and the stability of beaches and barrier systems \citep{ashton2001formation}. During extreme events such as tropical cyclones and extra-tropical storms, wave direction determines the spatial distribution of coastal impacts, shaping inundation patterns and infrastructure damage. Basin-scale analyses have shown that changes in wave climate, including shifts in dominant wave direction, are closely linked to climate variability and long-term ocean–atmosphere interactions \citep{young2011global}. More recently, satellite altimeter and reanalysis datasets have revealed systematic global changes in extreme wave conditions, with implications for coastal risk and marine operations \citep{young2019multiplatform}. Wave direction also plays a critical role in upper-ocean mixing and air-sea momentum exchange, influencing large-scale climate processes. With the rapid expansion of satellite remote sensing and global wave reanalysis products, directional wave fields are now observed continuously over entire ocean basins, making wave direction analysis central to coastal hazard assessment, offshore engineering design, maritime safety, and climate impact studies.

In this paper, we analyze wave direction data from the 2004 Indian Ocean Tsunami, which occurred on December 26, 2004, at 07:58:53 local time (UTC+7), and the epicenter was located off the west coast of northern Sumatra, Indonesia, along the Sunda subduction zone. The earthquake, with a magnitude exceeding 9.0, ruptured approximately 1,300 km of the plate boundary between the Indo-Australian and Eurasian plates. The spatial variability in tsunami wave approach angles significantly influenced run-up heights and localized destruction along the coasts of Indonesia, Sri Lanka, India, and Thailand. Tsunami waves propagated across the entire Indian Ocean basin, causing fatalities and widespread coastal damage in Somalia, Kenya, and Tanzania \citep{ishii2005extent}. Post-event assessments emphasize that wave-propagation pathways and coastal orientation amplified impacts in some regions \citep{lay2005great}. Understanding directional wave propagation is therefore essential for improving early warning systems, coastal hazard mapping, and evacuation planning in tsunami-prone and cyclone-exposed regions. Here, we explore data from the ERA5 global wave reanalysis, produced by the European Centre for Medium-Range Weather Forecasts under the Copernicus Climate Change Service, available at a $0.5^\circ \times 0.5^\circ$ resolution. Considering the entire Indian Ocean basin, stretched between 60$^\circ$S to 30$^\circ$N, and 20$^\circ$E to 147$^\circ$E, we obtain wave direction data at 33,845 grid cells.

Research on directional data spans more than six decades and has evolved from foundational theoretical developments to sophisticated modern modeling frameworks. Early advances in circular inference were made by \cite{watson1961goodness, watson1962goodness} and \cite{stephens1963random, stephens1970use}, who developed goodness-of-fit procedures and distributional theory for circular variables. Comprehensive treatments of directional statistics are available in the monographs \cite{fisher1995statistical} and \cite{mardia2000directional}, while \cite{fisher1995statistical} offer an extensive discussion with emphasis on circular methods and practical applications, even including spatial and spatiotemporal ones. With advances in computational statistics, particularly Markov chain Monte Carlo (MCMC) and expectation–maximization (EM) algorithms, directional data analysis has transitioned from primarily descriptive approaches to fully inferential modeling. Regression models for circular responses have been developed in both frequentist and Bayesian frameworks \citep{harrison1988development, damien1999full, kato2008circular}. Time series models for circular processes have also been proposed \citep{coles1998inference}.

In the context of linear spatial datasets, the Gaussian process (GP) remains the standard modeling framework in spatial geostatistics because of its strong theoretical foundation and tractable inference \citep{gelfand2016spatial}. Here, the finite-dimensional joint distributions depend only on the mean process and the covariance kernel, which makes the analysis tractable. The development of wrapped Gaussian process (WGP) models for spatial directional data began with the seminal work of \cite{jona2012spatial}, who introduced a coherent Bayesian framework for modeling wave direction fields by wrapping a Gaussian process onto the unit circle. Their construction enables spatial dependence to be modeled directly on the angular scale while retaining tractable inference through latent winding numbers. Subsequent researches expand this framework in several directions; \citet{mastrantonio2016spatio} extend WGP models to spatiotemporal settings with nonseparable covariance structures, while \citet{marques2022non} propose nonstationary wrapped Gaussian processes for spatially varying dependence. Joint modeling of linear and circular variables is addressed by \citet{wang2015joint}, allowing coherent analysis of wave heights and directions, and related projected Gaussian process approaches are developed by \citet{wang2014modeling} as an alternative construction for directional dependence. Broader reviews of advances in directional statistics, including spatial circular models, are provided by \citet{pewsey2021recent}. Recent methodological developments have explored circular and manifold-valued processes in machine learning contexts, including wrapped GP regression on Riemannian manifolds \citep{mallasto2018wrapped} and flexible normalizing flows on tori and spheres \citep{rezende2020normalizing}. Applications in environmental and energy sciences, such as wind and wave modeling \citep{arrieta2024spatially}, further highlight the need for scalable spatial directional models. Despite these advances, computational challenges persist for large spatial domains, motivating continued development of efficient wrapped spatial processes.

Inference for a dense GP requires repeated evaluation of the determinant and inverse of the covariance matrix, resulting in cubic computational complexity \citep{hazra2025exploring}. This computation becomes infeasible for large spatial datasets like ours. To address this limitation, numerous scalable approximations have been proposed, including kernel convolution methods \citep{higdon2002space}, low-rank basis function representations \citep{wikle1999dimension}, predictive process models \citep{banerjee2008gaussian}, spectral likelihood approximations \citep{fuentes2007approximate}, composite likelihood and conditional approximations \citep{vecchia1988estimation, stein2004approximating}, covariance tapering \citep{furrer2006covariance}, and Gaussian Markov random field (GMRF) formulations \citep{rue2005gaussian, lindgren2011explicit}. While the above scalable GP models have been explored in various settings, including spatial extremes \citep{hazra2021estimating, hazra2023bayesian, hazra2025efficient}, to our knowledge, no wrapped construction of such models for large spatial directional datasets exists in the literature. 

In this paper, we propose a wrapped GMRF (WGMRF, henceforth) model to analyze the large spatial directional datasets. The model exploits the finite truncation of the infinite sum representation of the wrapped normal distribution, as proposed in \cite{jona2012spatial}. Further, we build an approximation of the underlying isotropic Gaussian process using a GMRF, constructed using a stochastic partial differential equation (SPDE) representation of the corresponding Mat\'ern correlation function, motivated by \cite{lindgren2011explicit}. We further discuss some theoretical properties regarding the circular correlation and predictive distributions. The computation is based on MCMC, specifically a Metropolis-within-Gibbs sampling scheme. Exploiting the sparsity of the precision matrix of the underlying GMRF speeds up Gibbs sampling. The latent process also includes a nugget component, which further speeds up sampling. Through simulation studies, we compare the proposed method with a standard non-spatial wrapped normal distribution model and a low-rank WGP. The low-rank WGP is also our novel contribution; while it can model large spatial directional datasets, it is less elegant than the proposed model. In the data application, we implement the proposed model to obtain statistical inferences. Here, we also perform 10-fold cross-validation to compare the same three models.

The rest of the paper is organized as follows. In Section \ref{sec:data_eda}, we discuss the large spatial wave direction data and certain exploratory analyses suggesting the need for spatial modeling. Section \ref{sec:background} discusses the necessary background on univariate and multivariate wrapped normal distributions and WGP, along with the finite truncation strategy of \cite{jona2012spatial}. In Section \ref{sec:methodology}, we discuss the proposed WGMRF model, its properties, and the Bayesian computation. Here, we also describe the construction of the low-rank WGP model, which we treat as a competing model. Section \ref{sec:simulation} compares the proposed method with a non-spatial wrapped normal distribution model and the competing low-rank WGP model. In Section \ref{sec:application}, we apply the proposed method to the wave direction data over the Indian Ocean basin during the 2004 Indian Ocean Tsunami, along with comparing the proposed and competing models using a 10-fold cross-validation. Section \ref{sec:conclusion} concludes.

\section{Data description and exploratory analysis}
\label{sec:data_eda}

\begin{figure}[t]
    \centering
    \includegraphics[width=0.7\textwidth]{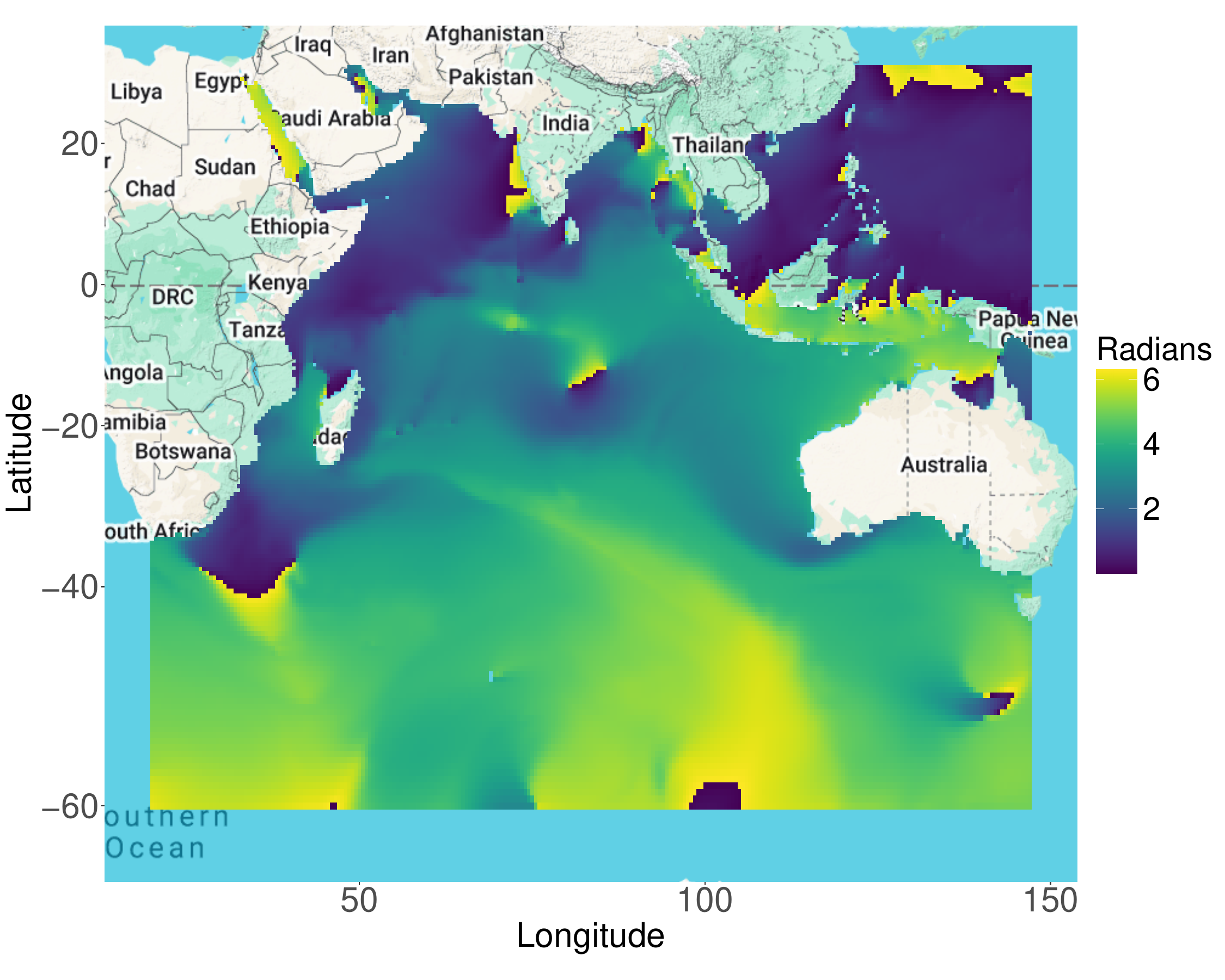}
    \caption{Gridded wave direction data obtained from the ERA5 global wave reanalysis, corresponding to the hour of the 2004 Indian Ocean Tsunami (01:00 hour UTC), covering the entire Indian Ocean basin.}
    \label{fig:wavedir_dataset}
\end{figure}

The wave direction data used in this study are obtained from the ERA5 global wave reanalysis produced by the European Centre for Medium-Range Weather Forecasts under the Copernicus Climate Change Service \citep{hersbach2020era5}. ERA5 provides globally consistent wave parameters derived from a coupled atmospheric–wave modeling system. The specific variable analyzed is the mean wave direction, defined as the average direction of ocean surface waves computed over all frequencies and directions of the two-dimensional wave spectrum. The wave spectrum represents the superposition of wind-sea waves, generated locally by winds, and swell waves, generated remotely and propagating across ocean basins \cite{holthuijsen2010waves}; the reported mean direction accounts for both components. The data are available on a regular latitude–longitude grid ($0.5^\circ \times 0.5^\circ$ resolution in this study) with hourly temporal resolution. Wave direction is reported in radians, indicating the direction from which waves are coming relative to geographic north (e.g., $0$ denotes waves coming from the north, $\pi/2$ from the east). While the dataset is available across all ocean basins, for our analysis we select the portion of the Indian Ocean basin, spanning 60$^\circ$S to 30$^\circ$N and 20$^\circ$E to 147$^\circ$E, for the hour 01:00 UTC, closest to the event time 07:58:53 local time (UTC+7). Overall, we obtain wave direction data for 33,845 grid cells. The heatmap of the dataset is presented in Figure \ref{fig:wavedir_dataset}. Compared with timestamps taken sufficiently before and after the event (not shown), the spatial profile appears smoother during the event, possibly because the tsunami affects the entire basin. In certain regions, specifically near the eastern coast of China and the southeastern coasts of South Africa, as well as a few others like Indonesia, the wave direction values for nearby pixels differ by $2\pi$ (the color changes between blue/dark and yellow/light). These observations indicate the need for directional data analysis rather than treating the data as a linear dataset; when the angles are concentrated near a single value, away from 0 and $2\pi$, treating circular data as linear is reasonable.

\begin{figure}[t]
    \centering
    \includegraphics[height=0.35\linewidth]{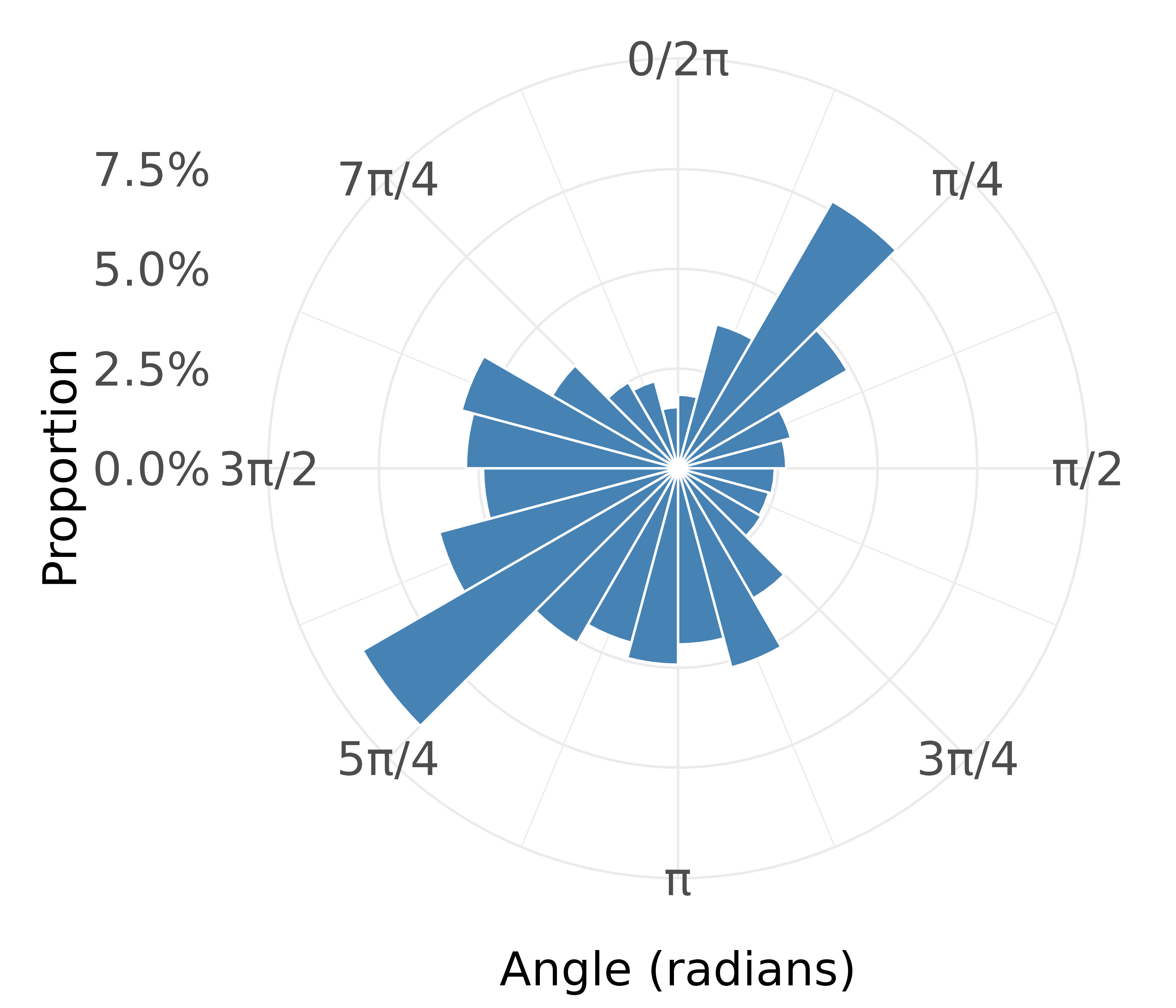}
    \includegraphics[height=0.35\linewidth]{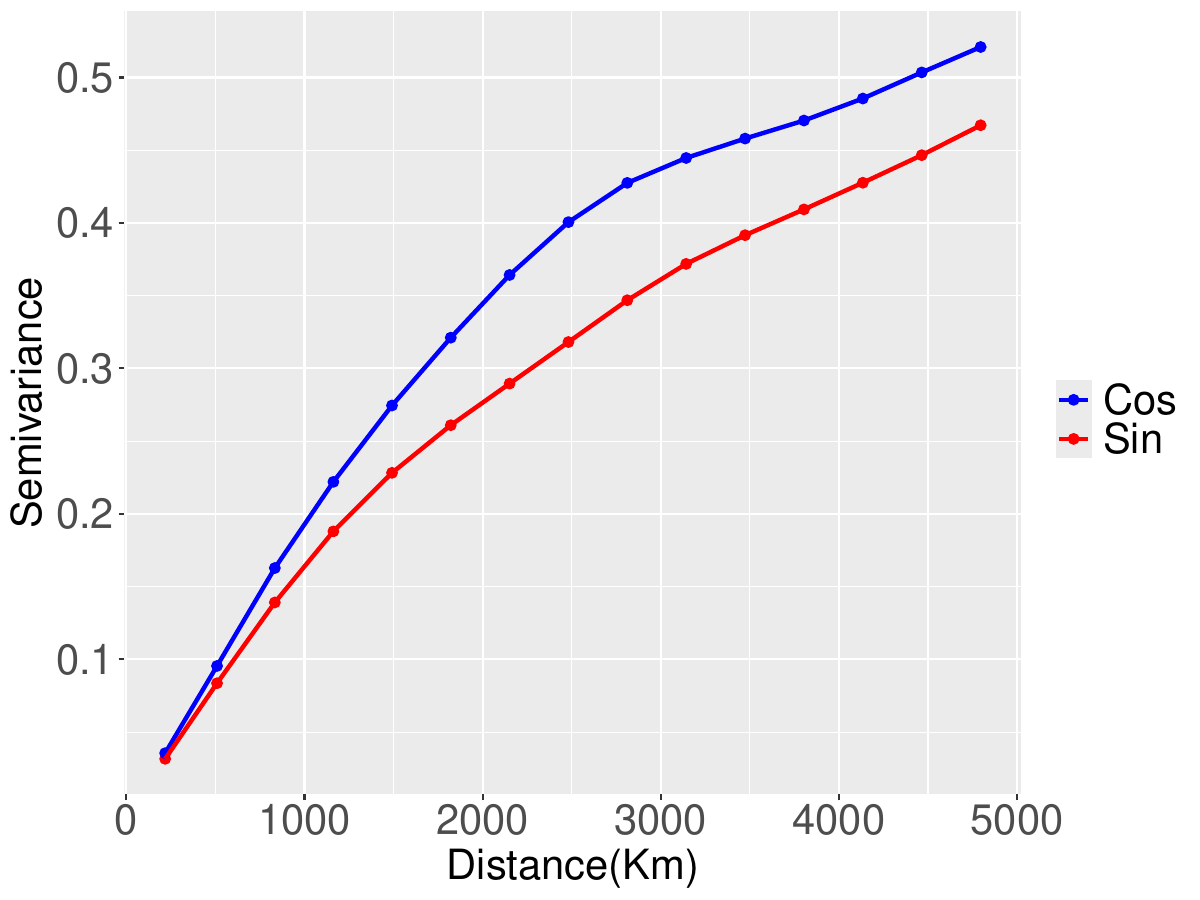}
    \caption{Left: Circular histogram of the wave direction dataset. Right: The empirical semivariograms of sine and cosine transformations of the wave direction dataset. Here, the $X$-axis denotes the geodesic distances.}
    \label{fig:eda}
\end{figure}

Before delving into a spatial analysis, we first explore the marginal data distribution, and we report the circular histogram in the left panel of Figure \ref{fig:eda}. The circular histogram reveals a pronounced bimodal structure in the angular data, with dominant modes centered around $\pi/4$ and $5\pi/4$. Notably, these two directions are separated by $\pi$, indicating that the field exhibits strong axial behavior, i.e., orientations aligned along a preferred northeast-southwest axis. In addition to these primary modes, relatively high proportions of observations are observed over the broader angular sector between $3\pi/4$ and $7\pi/4$, suggesting substantial directional variability around the southwest half-plane. This pattern indicates that, while the data are concentrated along a principal axis, there remains considerable dispersion, potentially exhibiting the influence of secondary physical mechanisms or localized perturbations. While the circular histogram shows a bimodal nature, strongly spatially correlated data from a unimodal density may induce patches of regions with two different angles.

To explore the spatial dependence, we present the empirical semivariograms of sine and cosine transformations of the wave direction data in the right panel of Figure \ref{fig:eda}. On the $X$-axis, we consider geodesic distance, instead of the standard choice of Euclidean distance. This choice is due to the large data domain and the spherical shape of the globe. Both exhibit clear spatial structure, with semivariance increasing with separation distance before approaching a sill, thereby providing strong evidence of spatial dependence in the directional field. The consistency of this pattern across both trigonometric components confirms that the spatial correlation is not an artifact of angular wrapping but reflects genuine dependence in the underlying process. Together, the bimodal directional distribution and the evident strong spatial correlation suggest that the angular field is governed by coherent large-scale dynamics with structured spatial organization, motivating the use of spatial directional models capable of capturing spatial dependence, even at large geographical distances. While wrapping a mixture of two GPs or GMRFs can be a more suitable model, for simplicity, we focus on wrapping a single GP or GMRF in this paper.

\section{Background}
\label{sec:background}

\subsection{Univariate wrapped distributions}

The von Mises distribution is the most common model for univariate circular data \citep{mardia1975statistics, mardia2000directional}. Although it is well-studied and computational tools for inference are well developed, its multivariate extensions remain challenging. Existing bivariate and trivariate generalizations are computationally demanding \citep{mardia2008multivariate}, making them less suitable even for moderately large spatial datasets.

An alternative construction is to wrap a linear random variable \citep{jona2012spatial}. 
Let $X \in \mathbb{R}$ be a linear (unwrapped, latent in our case) random variable with probability density function $f(x)$ and distribution function $F(x)$. The wrapped variable (response, in our case) $Y$ with period $2\pi$ is defined as $Y = X \ \mathrm{mod} \ 2\pi$ so that $0 \le Y < 2\pi$. The density of $Y$ is obtained by summing the linear density over all 
integer shifts of length $2\pi$. Writing $X = Y + 2\pi K$, where $K \in \mathbb{Z}$ is the \emph{winding number}, the wrapped density, i.e., the density of $Y$ is
\begin{equation}
g(y) = \sum_{k=-\infty}^{\infty} f(y + 2\pi k), 
\quad 0 \le y < 2\pi.
\end{equation}

Thus, the joint density of $(Y,K)$ is $p(y,k) = f(y + 2\pi k),~ y \in [0,2\pi), \ k \in \mathbb{Z}$. There is a one-to-one correspondence between $X$ and $(Y, K)$; each determines the other. Marginalizing over $K$ yields the wrapped density 
$g(y)$. The marginal distribution of $K$ is $\mathrm{P}(K = k) = \int_0^{2\pi} f(y + 2\pi k)\, dy.$ The conditional distributions of $K$ given $Y=y$ and $Y$ given $K=k$ are
\[
\mathrm{P}(K = k \mid Y = y)
=
\frac{f(y + 2\pi k)}
{\sum_{j=-\infty}^{\infty} f(y + 2\pi j)}, \quad \quad 
\pi(y \mid K = k)
=
\frac{f(y + 2\pi k)}
{\int_0^{2\pi} f(y + 2\pi k)\, dy},
\]
i.e., the conditional density of $Y$ given $K=k$ and the conditional probability mass function of $K$ given $Y=y$ are proportional to  $f(y + 2\pi k)$. These representations make wrapped models convenient for hierarchical and simulation-based inference, where $K$ can be treated as a latent variable. In practice, the infinite sum over $k$ is approximated by automatic truncation.

Direct computation of moments from $g(x)$ is generally difficult. Instead, it is convenient to work with the complex-valued variable $Z = e^{iY},$ which lies on the unit circle in the complex plane. For integer $p$, $\mathbb{E}\big(e^{ipY}\big) = \psi_X(p),$ where $\psi_X(\cdot)$ is the characteristic function of $X$, the linear variable. This result provides a simple way to obtain circular moments from the characteristic function of the underlying linear distribution.

\subsection{Wrapped univariate normal distribution}
\label{subsec:wn_univ}

The wrapped normal (WN) distribution is obtained by wrapping a normal random variable $X \sim \mathrm{N}(\mu, \sigma^2)$ onto the circle. We denote this by $Y \sim \text{WN}(\mu, \sigma^2)$. The mean parameter $\mu$ may be decomposed as $\mu = \tilde{\mu} + 2\pi K_\mu,$ where $\tilde{\mu} \in [0,2\pi)$ is the mean direction and $K_\mu$ is the corresponding winding number. The variance parameter $\sigma^2$ is often 
reparameterized as $c = e^{-\sigma^2/2}, \quad 0 < c < 1,$ where $c$ is called the concentration parameter.

The wrapped normal density is
\begin{equation}
g(y) =
\frac{1}{\sqrt{2\pi\sigma^2}}
\sum_{k=-\infty}^{\infty}
\exp\left\{
-\frac{(y - \mu + 2\pi k)^2}{2\sigma^2}
\right\},
\quad 0 \le y < 2\pi.
\end{equation}

Using the complex representation $Z = e^{iY}$, the $p$th circular moment is $E(Z^p) = e^{ip\mu - p^2\sigma^2/2}$. In particular, $\mathrm{E}(Z) = e^{-\sigma^2/2}(\cos\mu + i\sin\mu)$, so the resultant length equals $c = e^{-\sigma^2/2}$. Thus, $\mathrm{E}(\cos Y) = c\cos\mu$ and $\mathrm{E}(\sin Y) = c\sin\mu$ and the mean direction is $\tilde{\mu} = \mathrm{atan}^*(\mathrm{E}(\sin Y), \mathrm{E}(\cos Y))$.

\subsubsection*{Latent Variable Representation and Truncation}

For MCMC implementation, the winding number $K$ is introduced as a latent variable. Sampling $K$ is challenging because its support is $\mathbb{Z}$, but the infinite sum in the density can be accurately approximated by truncation. After translating to symmetric support $[-\pi,\pi)$, the wrapped density can be written as a sum of normal probabilities over intervals. Using the fact that approximately 99.7\% of normal probability mass lies 
within $\pm 3\sigma$, one can determine bounds $k_L$ and $k_U$ such that 
only terms with $k_L \le k \le k_U$ are needed. This result yields practical truncation rules:
\[
\begin{cases}
k \in \{-1,0,1\}, & \sigma < 2\pi/3, \\
k \in \{-2,-1,0,1,2\}, & 2\pi/3 \le \sigma < 4\pi/3,
\end{cases}
\]
and hence, large values of $K$ occur only when $\sigma^2$ is large. Thus, in our application, we set a practical choice $k \in \{-3, -2, \ldots, 2, 3\}$.

\subsubsection*{Identifiability and Practical Issues}

When $\sigma^2$ is large (equivalently, $c$ is small), the wrapped normal distribution approaches the uniform distribution on the circle. Consequently, $\sigma^2$ and $K$ may be poorly identified in Bayesian model fitting unless informative priors are imposed, as pointed out by \cite{jona2012spatial}. Simulation studies of \cite{jona2012spatial} show that standard uniformity tests (e.g., Rayleigh, Kuiper–Watson, Rao tests) may fail to distinguish WN from uniform data when $\sigma^2$ is sufficiently large, especially for small sample sizes. Therefore, exploratory analysis is recommended before fitting a WN model. Given observations $y_1,\dots,y_N$, define $\bar{C} = N^{-1}\sum_{i=1}^N \cos y_i$ and $\bar{S} = N^{-1}\sum_{i=1}^N \sin y_i$. Let $\hat{c} = \sqrt{\bar{C}^2 + \bar{S}^2}$. Then the moment estimators are $\hat{\tilde{\mu}} = \mathrm{atan}^*(\bar{S}, \bar{C})$ and $\hat{\sigma}^2 = -2\log \hat{c}$. These provide simple preliminary estimates for assessing the suitability of the wrapped normal model.

In our application, the obtained estimates are $\hat{\tilde{\mu}} = -2.2627$ and $\hat{\sigma} = 1.9220 < 2\pi / 3$. Thus, the above-mentioned limitation of the wrapped normal distribution does not apply to our case.

\subsubsection{Bayesian inference}

Suppose we observe independent and identically distributed angular data $y_1, y_2, \dots, y_N$ from a wrapped normal distribution. For Bayesian inference, it is convenient to work with the augmented representation involving the latent winding numbers $\{K_i\}_{i=1}^N$. The model is written in terms of the joint distribution of $\{(Y_i, K_i)\}_{i=1}^N$ given $(\mu, \sigma^2)$:
\[
p(\mathbf{y}, \mathbf{k} \mid \mu, \sigma^2)
=
\prod_{i=1}^N
\frac{1}{\sigma}
\phi\!\left(
\frac{y_i + 2\pi k_i - \mu}{\sigma}
\right),
\]
where $\phi(\cdot)$ is the standard normal density. Priors need to be assigned to $\mu$ and $\sigma^2$; specifically, given $\{K_i\}_{i=1}^N$, $X_i = Y_i + 2 \pi K_i$ for $i=1, \ldots, N$, and $X_i \overset{\mathrm{IID}}{\sim} \mathrm{N}(\mu, \sigma^2)$. Thus, it is possible to choose conjugate priors for $\mu$ and $\sigma^2$ (normal and inverse gamma, respectively). Overall, the posterior distribution is then constructed for $(\mu, \sigma^2, K_1, \dots, K_n)$. The latent variables $K_i$ are introduced solely to facilitate computation; posterior inference focuses on $\mu$, and hence the mean direction 
$\tilde{\mu}$, and $\sigma^2$.

In a Gibbs sampling scheme, each $K_i$ is updated conditional on $(\mu, \sigma^2, x_i)$. Although $K_i$ has support on $\mathbb{Z}$, the infinite support can be approximated using adaptive truncation. Using the normal tail bound (approximately 99.7\% mass within 
$\pm 3\sigma$), define $m = 1 + \frac{3\sigma}{2\pi},~k_i \in \{-m, \dots, 0, \dots, m\}.$ Then the conditional distribution of $K_i$ is approximated by
\begin{equation}\label{eq:k_update_iid}
\mathrm{P}(K_i = k \mid \mu, \sigma, y_i)
\approx
\frac{
\phi\!\left(\dfrac{y_i + 2\pi k - \mu}{\sigma}\right)
}{
\sum_{j=-m}^{m}
\phi\!\left(\dfrac{y_i + 2\pi j - \mu}{\sigma}\right)
},
\quad k=-m,\dots,m.
\end{equation}

This truncation ensures a uniformly small approximation error at each iteration while keeping computation efficient. However, in our application, we fix the lower and upper bounds for $K_i$ at -3 and 3, respectively, and the computation still remains scalable.

\subsection{Wrapped Gaussian processes}

In spatial applications, directional observations $\bm{Y}=(Y_1, \ldots, Y_N)^\top \in [0,2\pi)^N$ are often collected at $N$ different spatial locations, necessitating multivariate modeling with dependence. A general construction of a multivariate wrapped distribution includes a multivariate linear random vector 
$\bm{X} = (X_1, \ldots, X_N)^\top \in \mathbb{R}^N$ having density $f(\cdot;\bm{\theta})$ with $\bm{\theta}$ being the vector of model parameters. \cite{jona2012spatial} define $\bm{X} = \bm{Y} + 2\pi \bm{K}$ where $\bm{K} = (K_1, \ldots, K_N)^\top \in \mathbb{Z}^N$. The joint density of $(\bm{Y},\bm{K})$ is $p(\bm{y},\bm{k}) = f(\bm{y} + 2\pi \bm{k})$, and the marginal density of $\bm{Y}$ is obtained by summing over $\mathbb{Z}^n$. This results in an $N$-fold infinite sum, which is computationally intractable even for moderate $N$. As in the univariate case, introducing latent winding numbers $\bm{K} = (K_1, \ldots, K_N)^\top$ facilitates Bayesian computation.

When $f(\cdot;\bm{\theta})$ is multivariate normal with parameters $\bm{\theta} = (\bm{\mu}, \bm{\Sigma})$, we call $\bm{X} \sim \text{WN}_N(\bm{\mu}, \bm{\Sigma})$, an $N$-dimensional multivariate wrapped normal distribution. Because the conditional distributions of a multivariate normal are explicit, the corresponding conditional distributions for $(X_i, K_i)$ are straightforward to derive for Gibbs sampling.

A Gaussian process (GP) defined over the spatial domain of interest naturally induces a wrapped Gaussian process (WGP) through the wrapping transformation. Let $X(\cdot)$ be a Gaussian process indexed by spatial location $\bm{s} \in \mathcal{D} \subset \mathbb{R}^2$ with mean function $\mu(\cdot)$ and covariance function $\text{Cov}(X(\bm{s}), X(\bm{s}')) = \sigma^2 \rho(s - s'; \psi)$, where $\rho(\cdot;\psi)$ is a correlation function with decay parameter $\psi$. We can then define the wrapped process
\[
Y(\bm{s}) = X(\bm{s}) \ \text{mod } 2\pi.
\]

For a finite set of locations $\bm{s}_1, \ldots, \bm{s}_N$, the random vector $\bm{Y}$ follows a multivariate wrapped normal distribution $\bm{Y} \sim \text{WN}_N(\bm{\mu}, \sigma^2 \bm{R}(\phi)),$ where $\bm{\mu} = (\mu(\bm{s}_1), \ldots, \mu(\bm{s}_N))^\top,$ and $\bm{R}(\psi)$ is the correlation matrix with entries $R_{ij}(\psi) = \rho(s_i, s_j; \psi)$. Thus, the finite-dimensional distributions of the wrapped processes are obtained directly from those of the underlying GPs. In practice, one often assumes stationarity, i.e., $\rho(s_i, s_j; \psi) \equiv \rho(s_i - s_j; \psi)$ and isotropy, i.e., $\rho(s_i, s_j; \psi) \equiv \rho(\lVert s_i - s_j \rVert; \psi)$, although other covariance structures can be accommodated. Besides, unless some appropriate covariates are available, it is common to use $\mu(\bm{s}) = \mu$, constant over the spatial domain $\mathcal{D}$.

In the MCMC implementation of this method in \cite{jona2012spatial}, a key step lies in updating the latent $K_i$'s, similar to \eqref{eq:k_update_iid}. However, for their proposed WGP model, the numerator and denominator need to be calculated based on the conditional Gaussian densities rather than the simple univariate Gaussian densities in \eqref{eq:k_update_iid}. As a result, to calculate the conditional distribution of one component of an $N$-dimensional multivariate normal distribution, one needs to compute the inverse of an $(N-1) \times (N-1)$-dimensional matrix. Besides, we need to calculate it for each $K_i$ at each MCMC iteration, making it practically infeasible when $N$ is large.

Including a nugget component can significantly resolve this computational issue and has been explored in \cite{mastrantonio2016spatio}, although in a spatiotemporal context. Here, in our WGP setting, we can assume $\text{Cov}(X(\bm{s}), X(\bm{s}')) = \sigma^2[r \rho^*(s - s'; \phi) + (1-r) \delta(\bm{s}~=~\bm{s}')]$, where $\delta(\bm{s} = \bm{s}') = 1$ if $\bm{s} = \bm{s}'$ and zero otherwise. Thus, we can define $X(\bm{s}) = W(\bm{s}) + \varepsilon(\bm{s})$, where $W(\cdot)$ is a GP with mean function $\mu(\cdot)$ and covariance function $\text{Cov}(X(\bm{s}), X(\bm{s}')) = r\sigma^2 \rho^*(s - s'; \psi)$ and $\varepsilon(\bm{s}) \overset{\mathrm{IID}}{\sim} \mathrm{N}(0, (1-r)\sigma^2)$, IID over $\bm{s}$. Thus, for $\bm{W} = (W_1, \ldots, W_N)^\top$ with $W_i \equiv W(\bm{s}_i)$, the conditional distribution of $\bm{X}$ given $\bm{W}$ is multivariate normal has mean $\bm{W}$ and covariance matrix is $(1-r)\sigma^2 I_N$, where $I_N$ is the identity matrix. As a result, given another latent vector $\bm{W}$, the conditional distribution of $X_i$ does not depend on the other components, and thus, we can update $K_i$s in parallel.

Subsequently, when $N$ is large, even after including a nugget component, updating the spatially colored latent process $W(\cdot)$ is challenging, and we explore the SPDE-based construction \cite{lindgren2011explicit} in this context.

\section{Methodology}
\label{sec:methodology}

\subsection{SPDE-based Gaussian processes}
\label{subsec:spde_construction}

We first focus on constructing a standard (zero mean and unit variance) Gaussian process $Z(\cdot)$ having an isotropic Mat\'ern correlation function with nugget effect,
\begin{equation}
\label{cov_structure}
\rho(\bm{s}_1,\bm{s}_2)
=
\frac{r}{\Gamma(\nu) 2^{\nu-1}}
\left(
\frac{d(\bm{s}_1,\bm{s}_2)}{\psi}
\right)^{\nu}
K_\nu
\left(
\frac{d(\bm{s}_1,\bm{s}_2)}{\psi}
\right)
+
(1-r)\mathbb{I}(\bm{s}_1=\bm{s}_2),
\end{equation}
where $d(\bm{s}_1,\bm{s}_2)$ denotes geodesic distance (potentially scaled, not necessarily in kilometers), $\psi>0$ is the range parameter, $\nu>0$ controls smoothness, $r\in[0,1]$ represents the proportion of spatial variation, 
$K_\nu$ is the modified Bessel function of the second kind, and $\mathbb{I}(\cdot)$ is the indicator function. When $r=1$, integer values of $\nu$ determine the mean-square 
differentiability of $Z(\cdot)$. Since $\nu$ is typically weakly identified in practice \citep{cisneros2023combined, sahoo2025computationally}, it is often fixed; here we set $\nu=1$. The case of $\nu=2$ is explored in the supplementary material of \cite{hazra2025efficient}. However, the references explored the situation of $d(\bm{s}_1,\bm{s}_2)$ being Euclidean distance; in our case, because of the large geographical domain over the globe, such a choice is less practical.

Direct use of the Mat\'ern covariance in \eqref{cov_structure} leads to dense covariance matrices and high computational cost. To obtain a scalable representation, we approximate $Z(\cdot)$ with a GMRF defined on a finite mesh, exploiting the equivalence between Mat\'ern Gaussian processes and solutions to SPDEs \citep{lindgren2011explicit}, yielding an approximate process $\tilde{Z}(\cdot)$.

\begin{figure}[t]
    \centering
    \includegraphics[width=0.55\textwidth]{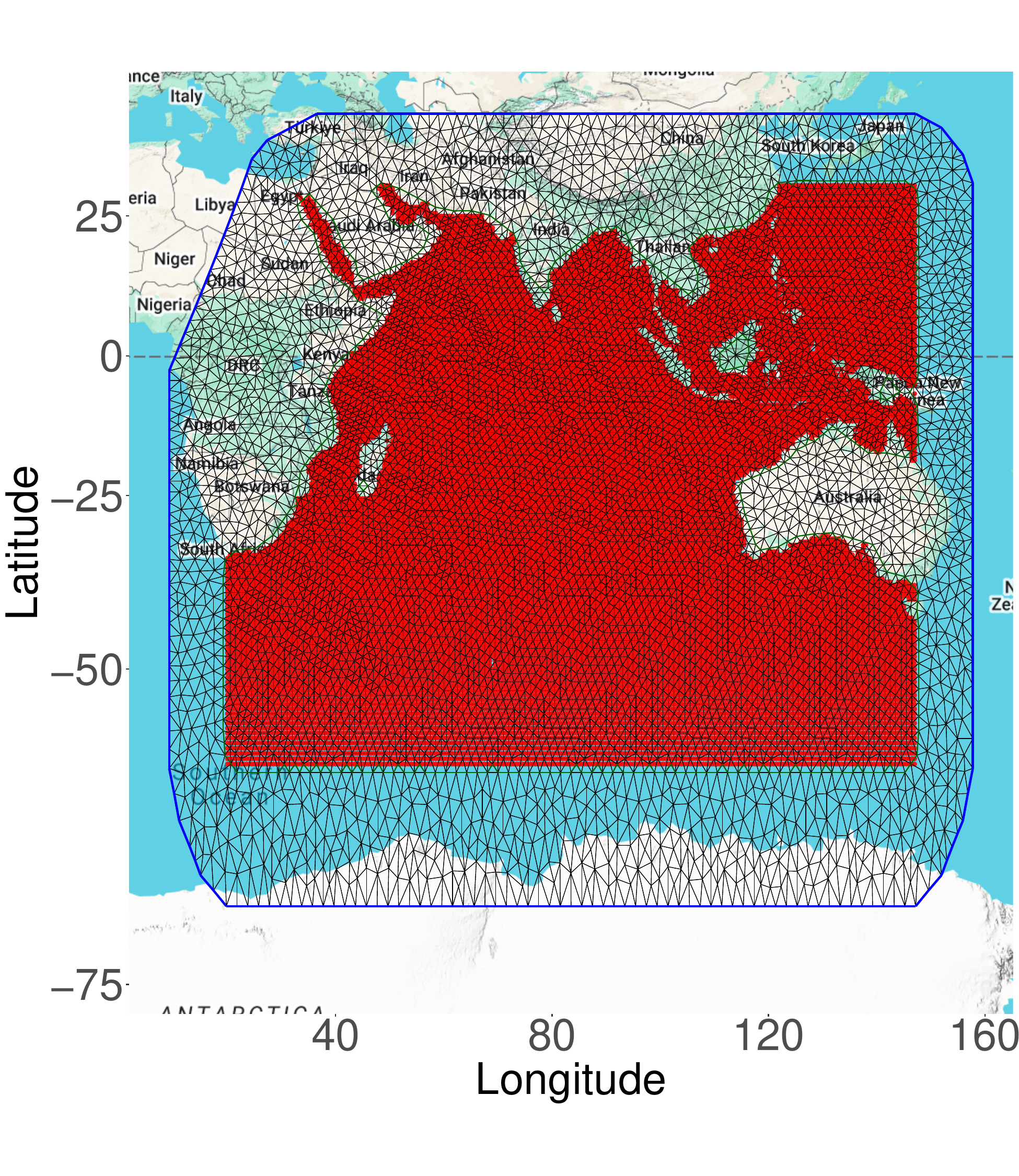}
    \caption{Triangulated mesh over the entire Indian Ocean basin, which we use to develop the spatial process $Z(\cdot)$ in \eqref{eq:spde}, using SPDEs. Given the spherical nature of the globe, we use \texttt{crs = inla.CRS("+proj=longlat +datum=WGS84")}. Here, the maximum triangle edge length near data and in outer extension are set to $1^\circ$ and $3^\circ$. The extension before the outer layer and for the coarsest triangles are set to $1^\circ$ and $10^\circ$, respectively. For choosing the boundary, we use a non-convex hull with the shrinkage amount $0.005^\circ$. Overall, the mesh includes 8145 nodes.}
    \label{fig:mesh_nodes}
\end{figure}

Let $\tilde\varepsilon(\cdot)$ denote a Gaussian process with correlation 
structure \eqref{cov_structure} and $r=1$. For $\nu=1$, 
$\tilde\varepsilon(\cdot)$ solves the SPDE
\begin{equation}
\label{eq:spde}
(\psi^{-2}-\nabla)\tilde\varepsilon(\bm{s})
=
4\pi\psi^{-2}\mathcal W(\bm{s}),
\qquad \bm{s}\in\mathbb R^2,
\end{equation}
where $\mathcal W(\bm{s})$ is Gaussian white noise and 
$\nabla$ is the Laplacian operator. 

We approximate the solution of \eqref{eq:spde} using finite element methods over a triangulated mesh covering $\mathcal D$. Let $\mathcal S=\{\bm{s}_1,\dots,\bm{s}_N\}$ denote observation locations and $\mathcal S^*=\{\bm{s}^*_1,\dots,\bm{s}^*_{N^*}\}$ denote the mesh nodes presented in Figure \ref{fig:mesh_nodes}. Here, $N^* = 8145$. Given the spherical nature of the globe, we use \texttt{crs = inla.CRS("+proj=longlat +datum=WGS84")} within \texttt{inla.mesh.2d} so that the approximated correlation is comparable with the case of geodesic distance, instead of Euclidean distance. The finite element approximation is $\tilde{\varepsilon}(\bm{s})~=~\sum_{j=1}^{N^*} \zeta_j(\bm{s})\varepsilon_j^*,$ where $\zeta_j(\cdot)$ are compactly supported, piecewise linear 
basis functions and $\bm{\varepsilon}^*
=
(\varepsilon_1^*,\dots,\varepsilon_{N^*}^*)^\top$ 
are Gaussian weights. It can be shown that $\bm{\varepsilon}^* \sim \text{Normal}_{N^*}(\bm 0,\bm Q_\psi^{-1}),$ with a sparse precision matrix
\begin{equation}\label{eq:precmat}
 \bm Q_\psi = \frac{1}{4\pi} \left( \psi^{-2}\bm D + 2\bm G_1 + \psi^{2}\bm G_2 \right),
\end{equation}
where $\bm D$, $\bm G_1$, and $\bm G_2$ are sparse finite-element matrices. Writing 
$\langle \tilde f, \tilde g\rangle=\int \tilde f(\bm s) \tilde g(\bm s)\,d\bm s$, 
the entries are $D_{jj}=\langle \zeta_j,1\rangle$, $(G_1)_{j_1j_2} =\langle \nabla\zeta_{j_1},\nabla\zeta_{j_2}\rangle$, $\bm G_2=\bm G_1\bm D^{-1}\bm G_1$. Here, $\text{Normal}_N$ denotes an $N$-variate normal distribution.

To evaluate the process at observation sites, we define the $(N\times N^*)$ projection matrix $\bm A$ with entries $A_{ij}=\zeta_j(\bm s_i)$, so that the projected field is 
$\bm A\bm{\varepsilon}^*$. Including the nugget effect, the approximate latent field 
$\tilde{\bm Z}
=
(\tilde Z(\bm s_1),\dots,\tilde Z(\bm s_N))^\top$ 
is
\begin{equation}
\label{z_construction_eq}
\tilde{\bm Z}
=
\sqrt r\,\bm A\bm{\varepsilon}^*
+
\sqrt{1-r}\,\bm\eta,
\end{equation}
where $\bm\eta\sim\text{Normal}_N(\bm 0,\bm I_N)$ independently. 
Its covariance matrix is
\begin{equation}\label{eq:approx_matern_cormat}
  \tilde{\bm{\Sigma}}_{\psi, r} = r\bm A\bm Q_\psi^{-1}\bm A^\top + (1-r)\bm I_N.
\end{equation}

A straightforward numerical comparison between the exact Mat\'ern correlation \eqref{cov_structure}, i.e., a correlation matrix $\bm{\Sigma}_{\psi, r}$ calculated using \eqref{cov_structure} at $\mathcal{S}$ and the SPDE-based approximation $\tilde{\bm{\Sigma}}_{\psi, r}$ in \eqref{eq:approx_matern_cormat} shows that the latter provides an accurate representation of the true spatial dependence \cite{lindgren2011explicit}. We therefore exploit the sparsity of $\bm Q_\psi$ for efficient Bayesian computation. As mentioned in the previous section, the conditional distribution
\[
\tilde{\bm Z}\mid \bm{\varepsilon}^*
\sim
\text{Normal}_N\big(
\sqrt r\,\bm A\bm{\varepsilon}^*,
(1-r)\bm I_N
\big)
\]
enables fast univariate imputation of $\tilde{\bm Z}$ and associated entities (like $K_i$'s in our wrapped model).

\subsection{Full model specification}

Here, we assume that the underlying Gaussian process has a constant marginal mean across its domain, i.e., $\mu(\bm{s}) = \mu$ for all $\bm{s} \in \mathcal{D}$. In our application, we do not have any meaningful covariate information available. However, if such information is available, proposing a regression model is straightforward. In the case of spatial regression with linear covariates, a monotone link function $h: \mathbb{R} \to (-\pi,\pi)$ with $h(0)=0$ (e.g., $h(z)=\arctan(z)$) can be used to map linear predictors to the circular scale. Finally, because of the constant mean assumption, we denote the mean vector by $\bm{\mu} = \mu(\bm{s}_1, \ldots, \mu(\bm{s}_N))^\top$ by $\mu \bm{1}_N$. 

Here, we describe the final SPDE-approximated model. In the data layer, we assume that the circular observations are $Y(\bm{s}_i) = X(\bm{s}_i) \mod 2\pi$ for $i=1,\ldots, N$. Given the location-specific winding numbers $K(\bm{s}_i)$'s, we have $X(\bm{s}_i) = Y(\bm{s}_i) + 2\pi K(\bm{s}_i)$. Here, both $X(\bm{s}_i)$ and $K(\bm{s}_i)$ are latent variables. Further, given the mesh-specific random effects vector $\tilde{\bm{\varepsilon}}^*$, the vector $\bm{X} = [X(\bm{s}_1), \ldots, X(\bm{s}_N)]^\top $ follows 
$\bm{X} \mid \tilde{\bm{\varepsilon}}^* \sim \text{Normal}_N\big(\mu \bm{1}_N + \bm A \tilde{\bm{\varepsilon}}^*, (1-r)\sigma^2\bm I_N \big)$. The latent vector $\tilde{\bm{\varepsilon}}^*$ follows $\tilde{\bm{\varepsilon}}^* \sim \text{Normal}_{N^*}(\bm 0, r \sigma^2\bm Q_\psi^{-1})$. Thus, after integrating out the latent $\tilde{\bm{\varepsilon}}^*$, we have $\bm{X} \sim \text{Normal}_N (\mu \bm{1}_N, \sigma^2 \tilde{\bm{\Sigma}}_{\psi, r})$. Thus, after wrapping $X(\bm{s}_i)$'s, we obtain $\bm{Y} = [Y(\bm{s}_1), \ldots, Y(\bm{s}_N)]^\top \sim \mathrm{WN}_N(\mu \bm{1}_N, \sigma^2 \tilde{\bm{\Sigma}}_{\psi, r})$. Further, we assign priors to $\mu$, $\sigma^2$, $\psi$, and $r$.

Overall, the proposed hierarchical Bayesian model is as follows:
\begin{eqnarray} \label{eq:final_model}
\text{\textbf{Data layer:}} \nonumber \\[4pt]
Y(\bm{s}_i) &=& X(\bm{s}_i) \bmod 2\pi, 
\qquad i = 1,\ldots,N, \nonumber \\
X(\bm{s}_i) &=& Y(\bm{s}_i) + 2\pi K(\bm{s}_i),
\qquad K(\bm{s}_i) \in \{-3,-2,\ldots,2,3\}, \nonumber
\\[8pt]
\text{\textbf{Process layer:}} \nonumber \\[4pt]
\bm{X} \mid \tilde{\bm{\varepsilon}}^*, \mu, \sigma^2, r 
&\sim& \mathrm{Normal}_N
\Big(
\mu \bm{1}_N + \bm{A}\tilde{\bm{\varepsilon}}^*,\,
(1-r)\sigma^2 \bm{I}_N
\Big), \nonumber \\
\tilde{\bm{\varepsilon}}^* \mid \sigma^2, r, \psi 
&\sim& \mathrm{Normal}_{N^*}
\Big(
\bm{0},\,
r\sigma^2 \bm{Q}_\psi^{-1}
\Big), \nonumber
\\[8pt]
\text{\textbf{Parameter layer:}} \nonumber \\[4pt]
[\mu, \sigma^2, \psi, r]  
&\sim& \pi(\mu, \sigma^2, \psi, r)
= \pi(\mu | \sigma^2)\,\pi(\sigma^2)\,\pi(\psi)\,\pi(r).
\end{eqnarray}

In our simulation studies and real-data application, we choose weakly informative priors $\mu | \sigma^2 \sim \mathrm{N}(0, 100^2 \sigma^2)$, $\sigma^2 \sim \mathrm{IG}(0.1, 0.1)$, $\psi \sim \mathrm{U}(0,\Delta)$, and $r \sim \mathrm{U}(0,1)$. Here, $\mathrm{IG}$ and $\mathrm{U}$ denote the inverse-gamma and uniform distributions, respectively. The notation $\Delta$ denotes the diameter of the study domain $\Delta = 155.66^\circ$ (in terms of Euclidean distance).

\subsection{Model properties}

\subsubsection{Correlation structure of the wrapped Gaussian process}

The WGP is defined by an underlying linear GP or linear GMRF, whose dependence structure is characterized by a covariance function and its associated parameters. In contrast to the linear setting, there is no unique definition of correlation for bivariate circular variables. A detailed discussion of appropriate circular correlation measures is given in 
\cite{jammalamadaka2001topics}. In particular, \cite{jammalamadaka1988correlation} proposed a circular correlation coefficient that satisfies many properties analogous to the classical product-moment correlation.

We consider a wrapped bivariate normal distribution for $(Y_1, Y_2)$ arising from a bivariate normal random vector with marginal variances $\sigma^2$ and correlation coefficient $\rho$. Under this construction, the circular correlation simplifies to $\rho_c(Y_1, Y_2) = \sinh(\rho \sigma^2) / \sinh(\sigma^2)$, and more generally, if the underlying Gaussian process has a covariance function $\text{Cov}(X(\bm{s}), X(\bm{s}')) = \sigma^2 \rho(\bm{s},\bm{s}'),$ the induced circular correlation function for the WGP is $\rho_c(s,s') = \sinh(\sigma^2 \rho(s,s')) / \sinh(\sigma^2).$

As pointed out by \cite{jona2012spatial}, although we do not construct covariance matrices directly using $\sinh(\rho(u))$, it is worth noting that this function defines a valid covariance function on $\mathbb{R}^d$ whenever $\sigma^2 \rho(u)$ is valid. Recalling the power series expansion $\sinh(v) = \sum_{n=0}^{\infty} \frac{v^{2n+1}}{(2n+1)!},$
if $\rho(u)$ is a valid correlation function on $\mathbb{R}^d$, then it is positive definite. A fundamental property of positive definite functions is that integer powers $\rho(u)^p$ remain positive definite for any integer $p \ge 1$. Therefore, $\sinh(\rho(u)) = \sum_{n=0}^{\infty} \frac{\rho(u)^{2n+1}}{(2n+1)!}$ is a nonnegative weighted sum of positive definite functions. Since nonnegative linear combinations of positive definite functions remain positive definite, $\sinh(\rho(u))$ is itself positive definite. Consequently, by the characterization of the stationary covariance functions via positive definiteness, $\sinh(\rho(u))$ defines a valid covariance function on $\mathbb{R}^d$.

In our setup, $\bm Q_\psi$ in \eqref{eq:precmat} is a positive definite matrix of dimension $N^* \times N^*$ (here, $N^*$ is the number of SPDE mesh nodes, which is smaller than $N$) and thus, its inverse $\bm Q_\psi$ is also positive definite. However, $\bm A\bm Q_\psi^{-1}\bm A^\top$ is a low-rank matrix of dimension $N \times N$ and thus, not positive definite. However, adding the nugget component, $\tilde{\bm{\Sigma}}_{\psi, r}$ is positive definite. Thus, if we consider any $(i,j)$th element of $\tilde{\bm{\Sigma}}_{\psi, r}$, say $\tilde{\bm{\Sigma}}^{(i,j)}_{\psi, r}$, it is appropriate to characterize the induced circular correlation between locations $\bm{s}_i$ and $\bm{s}_j$ by 
\begin{equation}\label{eq:cicular_cor_pairwise}
    \rho_c(\bm{s}_i, \bm{s}_j) = \dfrac{\sinh\big(\sigma^2 \tilde{\bm{\Sigma}}^{(i,j)}_{\psi, r}\big)}{\sinh(\sigma^2)}.
\end{equation}

\subsection{Bayesian computation}
\label{subsec:bayescomp}

We perform Bayesian inference using Markov chain Monte Carlo (MCMC), specifically Metropolis-within-Gibbs sampling. Here, the set of latent variables and parameters are $\bm{K} = [K(\bm{s}_1), \ldots, K(\bm{s}_N)]^\top$ (or equivalently, $\bm{X} = [X(\bm{s}_1), \ldots, X(\bm{s}_N)]^\top$), $\tilde{\bm{\varepsilon}}^*$, $\mu$, $\sigma^2$, $\psi$, and $r$. We draw MCMC samples from their posterior distribution given the data vector $\bm{Y} = [Y(\bm{s}_1), \ldots, Y(\bm{s}_N)]^\top$. We use the notation $\bm{A}(\bm{s}_i)$ to denote the $i$th row of $\bm{A}$; we treat it as a column vector. The Gibbs sampling steps are as follows.

\begin{itemize}
    \item The wrapping numbers $K(\bm{s}_1), \ldots, K(\bm{s}_N)$ are updated following the full conditional distribution (FCP) given by
    \begin{equation}
\nonumber \mathrm{P}(K(\bm{s}_i) = k \mid \mu, \sigma^2, r, \tilde{\bm{\varepsilon}}^*, \bm{Y}) = 
\frac{
\phi\!\left(\dfrac{Y(\bm{s}_i) + 2\pi k - \mu - \bm{A}(\bm{s}_i)^\top \tilde{\bm{\varepsilon}}^*}{\sigma \sqrt{1 - r}}\right)
}{
\sum_{j=-3}^{3}
\phi\!\left(\dfrac{Y(\bm{s}_i) + 2\pi j - \mu - \bm{A}(\bm{s}_i)^\top \tilde{\bm{\varepsilon}}^*}{\sigma \sqrt{1 - r}}\right)
},
\quad k=-3,\ldots,3,
\end{equation}
where $\phi(\cdot)$ denotes the standard normal density. Here, the update step of $K(\bm{s}_i)$ does not depend on $K(\bm{s}_j)$ for $j \neq i$. Thus, all the $K(\bm{s}_i)$'s can be updated in parallel, drastically reducing the computation time. Once a sample from $\bm{K}$, we equivalently obtain a sample from $\bm{X}$, which is used in further steps.

\item The FCP of $\tilde{\bm{\varepsilon}}^{*}$ is $\tilde{\bm{\varepsilon}}^{*} \mid\text{rest} \sim \textrm{Normal}_{N^*}(\bm{\mu}^\ast_{\varepsilon}, \bm{\Sigma}^\ast_{\varepsilon})$, where
\begin{eqnarray}
\nonumber && \bm{\Sigma}^\ast_{\varepsilon} = \sigma^2 \left[(1 - r)^{-1} \bm{A}'\bm{A} + r^{-1} \bm{Q}_{\psi} \right]^{-1}, ~~ \bm{\mu}^\ast_{\varepsilon} =  \bm{\Sigma}^\ast_{\varepsilon} (1 - r)^{-1} \bm{A}' [\bm{X} - \mu \bm{1}_N].
\end{eqnarray}

\item The FCP of $\mu$ is 
\begin{equation}
    \nonumber \mu \mid\text{rest} \sim \textrm{N}\left(\frac{\bm{1}'_N (\bm{X} - \bm{A} \tilde{\bm{\varepsilon}}^*) / (1 - r)}{N/(1-r) + 1/100^2}, \frac{\sigma^2}{N/(1-r) + 1/100^2}\right).
\end{equation}

\item The FCP of $\sigma^2$ is
\begin{eqnarray}
\nonumber   \sigma^2 \mid\text{rest} &\sim& \textrm{IG}\left(0.1 + (N + N^* + 1)/2, 0.1 + (r^{-1}\tilde{\bm{\varepsilon}}^{*'} \bm{Q}_{\psi} \tilde{\bm{\varepsilon}}^* \right. \\ \nonumber && \quad \quad \left.  + (1 - r)^{-1} ({\bm{X} - \mu \bm{1}_N  - \bm{A} \tilde{\bm{\varepsilon}}^*})' ({\bm{X} - \mu \bm{1}_N  - \bm{A} \tilde{\bm{\varepsilon}}^*}) + \mu^2)/2 \right).
\end{eqnarray}

\item We next discuss the update step for $\psi$. Suppose $\psi^{(m)}$ denotes the $m$-th MCMC sample from $\psi$. Considering a logit transformation, we obtain $\psi^{*(m)} \in \mathbb{R}$ from $\psi^{(m)}$, and simulate $\psi^{*(c)} \sim \textrm{Normal}( \psi^{*(m)}, s_{\psi}^2)$, where $s_{\psi}$ is the standard deviation of the candidate normal distribution. Subsequently, using an inverse-logit transformation, we obtain $\psi^{(c)}$ from $\psi^{*(c)}$, and consider $\psi^{(c)}$ to be a candidate from the posterior distribution of $\psi$. The acceptance ratio is 
\begin{eqnarray}
\nonumber \mathcal{R} &=& \frac{  f_{\textrm{Normal}_{N^*}}\left(\tilde{\bm{\varepsilon}}^*; \bm{0}_{N^*}, \sigma^2 r \bm{Q}_{\psi^{(c)}}^{-1} \right) }{   f_{\textrm{Normal}_{N^*}}\left(\tilde{\bm{\varepsilon}}^*; \bm{0}_{N^*}, \sigma^2 r \bm{Q}_{\psi^{(m)}}^{-1} \right)} \times \frac{\psi^{(c)} \left(\Delta - \psi^{(c)}\right)}{\psi^{(m)} \left(\Delta - \psi^{(m)}\right)},
\end{eqnarray}
where $f_{\textrm{Normal}_n}(\cdot; \bm{\mu}, \bm{\Sigma})$ denotes the $n$-variate normal density with mean $\bm{\mu}$ and covariance matrix $\bm{\Sigma}$. The candidate is accepted with probability $\min \lbrace \mathcal{R},1 \rbrace$.

\item Further, we discuss the update step of $r$. Suppose $r^{(m)}$ denotes the $m$-th MCMC sample from $r$. We simulate a candidate sample $r^{(c)}$ from $r^{(m)}$ following a procedure similar to simulating $\psi^{(c)}$ from $\psi^{(m)}$. The Metropolis-Hastings acceptance ratio is 
\begin{eqnarray}
\nonumber \mathcal{R} &=& \frac{ f_{\textrm{Normal}_N}\left(\bm{X}; \mu \bm{1}_N + \bm{A} \tilde{\bm{\varepsilon}}^*, \sigma^2 (1 - r^{(c)}) \bm{I}_N \right)}{  f_{\textrm{Normal}_N}\left(\bm{X}; \mu \bm{1}_N + \bm{A} \tilde{\bm{\varepsilon}}^*, \sigma^2 (1 - r^{(m)}) \bm{I}_N \right)} \\
\nonumber && \times \frac{  f_{\textrm{Normal}_{N^*}}\left(\tilde{\bm{\varepsilon}}^*; \bm{0}_{N^*}, \sigma^2 r^{(c)} \bm{Q}_{\psi}^{-1} \right) }{  f_{\textrm{Normal}_{N^*}}\left(\tilde{\bm{\varepsilon}}^*; \bm{0}_{N^*}, \sigma^2 r^{(m)} \bm{Q}_{\psi}^{-1} \right)} \times \frac{r^{(c)}  \left(1 - r^{(c)}\right)}{r^{(m)} \left(1 - r^{(m)}\right)}.
\end{eqnarray}
\end{itemize}

We implement the computation using \texttt{R} (\url{http://www.r-project.org}). For our main data application, we run the chain for 60,000 iterations, with the first 10,000 iterations as a burn-in period. Further, of the remaining 50,000 samples, we keep 10,000 after thinning the chains by a length of 5. The computation time on an AMD Ryzen 9 5900X processor with 64GB RAM is 316.16 minutes. For the simulation studies with 100 replications and the 10-fold cross-validation on the real dataset, we reduce the chain length; here, we draw 20,000 samples in each case, discard the first 10,000 as burn-in, and, after thinning by 5, draw inferences from the remaining 2,000 post-burn-in samples. The convergence and mixing are monitored via trace plots. We tune the candidate distributions within the burn-in phase to ensure the acceptance rate for Metropolis-Hastings steps remains between 0.3 and 0.5.

\subsection{Spatial prediction}

Spatial prediction under the proposed model proceeds analogously to kriging in GP models, but exploits the sparse precision structure of the underlying GMRF. Apart from the usual differences between the spatial prediction for a dense GP and a GMRF, the additional requirements are similar, and we follow the strategy of \cite{jona2012spatial}. 

Suppose we are interested in predicting the circular observation $Y(\bm{s}_0)$ at a new location $\bm{s}_0$. Given the random effects $\bm \tilde{\bm{\varepsilon}}^*$ and the parameters, the latent field at observation sites $\bm X$ follows $\bm{X} \mid \tilde{\bm{\varepsilon}}^*, \mu, \sigma^2, r 
\sim \mathrm{Normal}_N(\mu \bm{1}_N + \bm{A}\tilde{\bm{\varepsilon}}^*,\, (1-r)\sigma^2 \bm{I}_N)$. For prediction at $\bm{s}_0$, suppose the SPDE projection vector from the mesh nodes to $\bm{s}_0$ is denoted by $\bm{a}_0$. Thus, 
\begin{equation}\label{eq:spatial_pred}
    X(\bm{s}_0) \mid \tilde{\bm{\varepsilon}}^*, \mu, \sigma^2, r \sim \mathrm{N}(\mu + \bm{a}_0^\top\tilde{\bm{\varepsilon}}^*,\, (1-r)\sigma^2 \bm{I}_N), \quad Y(\bm{s}_0) = X(\bm{s}_0)~\mathrm{mod}~ 2\pi,
\end{equation}
where $Y(\bm{s}_0)$ is the wrapped variable corresponding to $X(\bm{s}_0)$. While we can draw posterior predictive samples from $Y(\bm{s}_0)$ using \eqref{eq:spatial_pred}, providing summary statistics is challenging, as simple averaging of the posterior samples from $Y(\bm{s}_0)$ does not provide the expectation of the posterior predictive distribution, due to the circular nature of the samples. As a result, we perform spatial prediction using circular moments. Here, the conditional wrapped moment is $$ \mathrm{E}\left[e^{iY(\bm s_0)} \mid \tilde{\bm{\varepsilon}}^*, \mu, \sigma^2, r \right] =
\exp\!\left(-\frac{(1-r)\sigma^2}{2} + i\big(\mu + \bm a_0^\top \tilde{\bm{\varepsilon}}^*\big)\right).$$

Within the Bayesian framework, posterior samples $\{ \tilde{\bm{\varepsilon}}^{*(b)}, \mu^{(b)}, \sigma^{2(b)}, r^{(b)} \}_{b=1}^B$ are obtained from MCMC. Monte Carlo integration yields
\[
\mathrm{E} \!\left[e^{i Y(\bm s_0)} \mid \bm{Y} \right] \approx \frac{1}{B} \sum_{b=1}^B
\exp\!\left(-\frac{(1-r^{(b)})\sigma^{2(b)}}{2} + i\big(\mu^{(b)} + \bm a_0^\top \tilde{\bm{\varepsilon}}^{*(b)}\big)\right).
\]

We define 
\begin{eqnarray}
\nonumber  g_c(\bm s_0) &=& \frac{1}{B} \sum_{b=1}^B \exp\!\left(-\frac{(1-r^{(b)})\sigma^{2(b)}}{2}\right) \cos\big(\mu^{(b)} + \bm a_0^\top \tilde{\bm{\varepsilon}}^{*(b)}\big), \\
\nonumber \quad g_s(\bm s_0)
&=& \frac{1}{B} \sum_{b=1}^B \exp\!\left(-\frac{(1-r^{(b)})\sigma^{2(b)}}{2}\right) \sin\big(\mu^{(b)} + \bm a_0^\top \tilde{\bm{\varepsilon}}^{*(b)}\big).
\end{eqnarray}

Finally, the posterior predictive mean direction and the associated posterior concentration, respectively, are
\begin{equation}\label{eq:krig_summary}
    \widehat{m}(\bm s_0) = \operatorname{atan*}\big(g_s(\bm s_0), g_c(\bm s_0)\big), \quad \widehat{c}(\bm s_0) = \sqrt{g_c(\bm s_0)^2 +g_s(\bm s_0)^2}.
\end{equation}
For a single-number predictive summary, we treat the predicted angle at $\bm s_0$ as $\widehat{Y}(\bm{s}_0) = \widehat{m}(\bm s_0)$. For comparing prediction uncertainty between models, we compare $\widehat{c}(\bm s_0)$.
\section{Simulation studies}
\label{sec:simulation}

Here, our main aim is to compare the proposed methodology with a non-spatial wrapped normal distribution model in Section \ref{sec:background}. In addition, we also compare our method with a simpler spatial model, namely a low-rank WGP, constructed based on the idea of \cite{wikle1999dimension}. We first describe the competing low-rank WGP model below, as it is our novel contribution and can improve model fitting over available methods suitable for a dataset like ours.

\subsection{Low-rank wrapped Gaussian processes}
\label{subsec:lowrank_wgp}

Let $\mathcal{S} = \{\bm s_1,\ldots,\bm s_N\}$ denote observation locations as before and
$\tilde{\mathcal{S}} = \{\tilde{\bm{s}}_1,\ldots,\tilde{\bm{s}}_M\}$ be a set of spatial knots with $M \ll N$. We approximate the latent GP as
\begin{equation}\label{eq:wrapped_lowrank}
    X(\bm s) = \mu + \bm b_{\phi}(\bm s)^\top \bm W + \varepsilon(\bm s),
\end{equation}
where $\mu \in \mathbb R$ is a constant mean, $\bm b_{\phi}(\bm s) = \big( C(\bm s, \tilde{\bm s}_1;\phi),\ldots, C(\bm s, \tilde{\bm s}_M;\phi) \big)^\top$ is a vector of basis functions
constructed from a correlation kernel $C(\cdot,\cdot;\phi)$ with bandwidth (range) parameter $\phi>0$, $\bm{W} = [W(\tilde{\bm{s}}_1), \ldots, W(\tilde{\bm{s}}_M)]^\top$, where we assume that $W(\tilde{\bm{s}}_i) \overset{\mathrm{IID}}{\sim} \mathrm{N}(0, \sigma^2)$, and $\varepsilon(\bm s) \overset{\mathrm{IID}}{\sim} \mathrm{N}(0,\tau^2)$ is a nugget term. In the literature, assuming a spatial correlation structure for $\bm{W}$ is also common. However, in such cases, the bandwidth $\phi$ is generally fixed. Here, prefer to keep the $W(\tilde{\bm{s}}_i)$'s uncorrelated, and consider the bandwidth $\phi$ as an unknown parameter. While other forms are also common, we choose $C(\bm s_i,\bm s_j^*;\phi) = \frac{1}{\sqrt{2\pi \phi^2}} \exp[-0.5\Vert \bm s_i -\bm s_j^* \Vert^2 / \phi^2]$, where $\Vert \bm s_i -\bm s_j^* \Vert$ denotes the Euclidean distance between $\bm s_i$ and $\bm s_j^*$.

For $\bm X = (X(\bm s_1),\ldots,X(\bm s_N))^\top$,
this yields $\bm X \mid \bm W \sim
\mathrm{Normal}_N
\big(\mu\bm 1_N + \bm B_{\phi} \bm W, \tau^2 \bm I_N \big),$
where $\bm B_{\phi}$ is the $N\times M$ matrix with
$(i,j)$th entry $C(\bm s_i,\bm s_j^*;\phi)$.
The resulting marginal covariance of $\bm X$ is $\bm \Sigma_Y = \sigma^2 \bm B_{\phi} \bm B_{\phi}^\top + \tau^2 \bm I_n$ provide a computationally efficient rank-$M$ approximation
to a full Gaussian process. Further, by wrapping $X(\bm{s})$ as $Y(\bm{s}) = X(\bm{s}) \mod 2\pi$, we obtain that $\bm Y = (Y(\bm s_1),\ldots,Y(\bm s_N))^\top$ follows $\bm Y \sim \mathrm{WN}_N(\mu\bm 1_N, \sigma^2 \bm B_{\phi} \bm B_{\phi}^\top + \tau^2 \bm I_n)$. As a result, the final model has circular spatial correlation, with reasonable flexibility, and serves as an alternative to our proposed methodology.

Introducing latent winding numbers $K(\bm{s}_i) \in \mathbb Z$ such that $X(\bm s_i) = Y(\bm s_i) + 2\pi K(\bm s_i),$ the joint density of $(Y(\bm{s}_i),K(\bm{s}_i))$ is Gaussian in $Y(\bm s_i) + 2\pi K(\bm s_i)$. We again truncate $K(\bm{s}_i)$ between -3 and 3. The likelihood is therefore written in terms of $\bm Y = \bm X + 2\pi \bm K$, with $\bm{K}=[K(\bm{s}_1, \ldots, K(\bm{s}_N)]^\top$. We further consider priors for $\mu$, $\sigma^2$, and $\tau^2$. We choose independent weakly informative conjugate priors $\mu \sim \mathrm{N}(0, 100^2)$, $\sigma^2 \sim \mathrm{IG}(0.1, 0.1)$, $\tau^2 \sim \mathrm{IG}(0.1, 0.1)$, and $\phi \sim \mathrm{U}(0, 0.2 \Delta)$, where $\Delta$ is the spatial domain diameter. In our simulation studies and application, we fix $M=100$. For Bayesian inference, we implement a Metropolis-within-Gibbs sampling algorithm, where we draw posterior samples from $\bm{K}$, $\bm{W}$, $\mu$, $\sigma^2$, and $\tau^2$.

\subsubsection{Posterior computation}
We draw MCMC samples from their posterior distribution given the data vector $\bm{Y} = [Y(\bm{s}_1), \ldots, Y(\bm{s}_N)]^\top$. The Gibbs sampling steps are as follows.

\begin{itemize}
    \item The wrapping numbers $K(\bm{s}_1), \ldots, K(\bm{s}_N)$ are updated following the full conditional distribution (FCP) given by
    \begin{equation}
\nonumber \mathrm{P}(K(\bm{s}_i) = k \mid \mu, \tau^2, \bm{W}, \bm{Y}) = 
\frac{
\phi\!\left([Y(\bm{s}_i) + 2\pi k - \mu - \bm{b}_{\phi}(\bm{s}_i)^\top \bm{W}] / \tau
\right)}{
\sum_{j=-3}^{3}
\phi\!\left([Y(\bm{s}_i) + 2\pi j - \mu - \bm{b}_{\phi}(\bm{s}_i)^\top \bm{W}] / \tau\right)
},
\quad k=-3,\ldots,3,
\end{equation}
where $\phi(\cdot)$ denotes the standard normal density. Here, similar to the proposed model, the update step of $K(\bm{s}_i)$ does not depend on $K(\bm{s}_j)$ for $j \neq i$. Thus, all the $K(\bm{s}_i)$'s can be updated in parallel as well. Once a sample from $\bm{K}$, we equivalently obtain a sample from $\bm{X}$, which is used in further steps.

\item The FCP of $\bm{W}$ is $\bm{W} \mid\text{rest} \sim \textrm{Normal}_{M}(\bm{\mu}_{W}, \bm{\Sigma}_{W})$, where $\bm{\Sigma}_{W} = \left[\tau^{-2} \bm{B}_{\phi}'\bm{B}_{\phi} + \sigma^{-2} \bm{I}_M \right]^{-1}$ and $\bm{\mu}_{W} = \tau^{-2} \bm{\Sigma}_{W}  \bm{B}_{\phi}' [\bm{X} - \mu \bm{1}_N]$.

\item The FCP of $\mu$ is $\mu \mid\text{rest} \sim \textrm{N}\left([\bm{1}'_N (\bm{X} - \bm{B}_{\phi} \bm{W}) / \tau^2]/[N/\tau^2 + 1/100^2], 1/[N/\tau^2 + 1/100^2]\right)$.

\item The FCP of $\sigma^2$ is $\sigma^2 \mid\text{rest} \sim \textrm{IG}\left(0.1 + M/2, 0.1 + \bm{W}'\bm{W} /2 \right).$

\item The FCP of $\tau^2$ is $\tau^2 \mid\text{rest} \sim \textrm{IG}\left(0.1 + N/2, 0.1 + (\bm{X} - \mu \bm{1}_N  - \bm{B}_{\phi} \bm{W})' (\bm{X} - \mu \bm{1}_N  - \bm{B}_{\phi} \bm{W})/2 \right)$.

\item The update step for $\phi$ is as follows. Suppose $\phi^{(m)}$ denotes the $m$-th MCMC sample from $\phi$. Considering a logit transformation, we obtain $\phi^{*(m)} \in \mathbb{R}$ from $\phi^{(m)}$, and simulate $\phi^{*(c)} \sim \textrm{Normal}( \phi^{*(m)}, s_{\phi}^2)$. Subsequently, using an inverse-logit transformation, we obtain $\phi^{(c)}$ from $\phi^{*(c)}$, and consider $\phi^{(c)}$ to be a candidate from the posterior distribution of $\phi$. The acceptance ratio is 
\begin{eqnarray}
\nonumber \mathcal{R} &=& \frac{  f_{\textrm{Normal}_{N}}\left(\bm{X}; \mu\bm 1_N + \bm B_{\phi^{(c)}} \bm W, \tau^2 \bm I_N \right) }{   f_{\textrm{Normal}_{N}}\left(\bm{X}; \mu\bm 1_N + \bm B_{\phi^{(m)}} \bm W, \tau^2 \bm I_N \right)} \times \frac{\phi^{(c)} \left(0.2\Delta - \phi^{(c)}\right)}{\phi^{(m)} \left(0.2\Delta - \phi^{(m)}\right)}.
\end{eqnarray}
\end{itemize}

The implementation, the choice of burn-in and post-burn-in MCMC sample sizes, and tuning of the candidate distribution are similar to those for the proposed WGMRF model. The computation time for the wrapped low-rank GP model is comparable to that of the proposed WGMRF model. While the low-rank method requires inverting an $M\times M$ matrix with $M=100$, the underlying matrix is not sparse. On the other hand, in the proposed WGMRF model, even with $N^* = 8145$ mesh nodes, the update step remains fast by exploiting the sparsity of the underlying GMRF.

\subsection{Validation metrics}
\label{subsec:validation_metrics}

We first divide the set of observation locations $\mathcal{S}$ into disjoint training and testing sets $\mathcal{S}_{\mathrm{train}}$ and $\mathcal{S}_{\mathrm{test}}$. Further, within the testing set, suppose the observed angles are $Y(\bm{s}_i)$'s where $\bm{s}_i \in \mathcal{S}_{\mathrm{test}}$. Let the corresponding predicted values based on fitting the proposed or a competing model be denoted by $\widehat{Y}(\bm{s}_i)$'s where $\bm{s}_i \in \mathcal{S}_{\mathrm{test}}$. Since angular variables lie on the unit circle, standard linear error measures are inappropriate because of the $2\pi$ wraparound. We therefore consider the following circularly valid comparison metrics. Here, we use the notation $\vert \mathcal{S}_{\mathrm{test}} \vert$ for cardinality of the set $\mathcal{S}_{\mathrm{test}}$.

\paragraph{The sine-cosine root mean squared error (SC-RMSE)} The (SC-RMSE) is defined as
\[
\mathrm{SC\text{-}RMSE} =
\sqrt{
\frac{1}{\vert \mathcal{S}_{\mathrm{test}} \vert} \sum_{\bm{s}_i \in \mathcal{S}_{\mathrm{test}}}\left\lbrace \left(\cos[Y(\bm{s}_i)] - \cos[\widehat{Y}(\bm{s}_i)]\right)^2 + \left(\sin[Y(\bm{s}_i)] - \sin[\widehat{Y}(\bm{s}_i)]\right)^2 \right\rbrace
}.
\]
A method providing a smaller SC-RMSE is preferable.

\paragraph{Circular Errors (Shortest Angular Distance)}
We define the circular difference as $\Delta(\bm{s}_{i}) = \operatorname{atan2}\big(\sin(\widehat{Y}(\bm{s}_i) - Y(\bm{s}_i)), \cos(\widehat{Y}(\bm{s}_i) - Y(\bm{s}_i))\big)$, which maps the discrepancy to $(-\pi,\pi]$. Based on this, we define the circular root mean squared error (CRMSE) and circular mean absolute error (CMAE) as
\[
\mathrm{CRMSE} = \sqrt{ \frac{1}{\vert \mathcal{S}_{\mathrm{test}} \vert}
\sum_{\bm{s}_i \in \mathcal{S}_{\mathrm{test}}}
\Delta(\bm{s}_{i})^2
}, \quad
\mathrm{CMAE}
=
\frac{1}{\vert \mathcal{S}_{\mathrm{test}} \vert}
\sum_{\bm{s}_i \in \mathcal{S}_{\mathrm{test}}}
|\Delta(\bm{s}_{i})|.
\]
A method providing a smaller CRMSE or CMAE is preferable.

\paragraph{Mean Resultant Length of Errors}

Expressing the circular errors in complex form,
\[
R_e = \left|
\frac{1}{\vert \mathcal{S}_{\mathrm{test}} \vert}
\sum_{\bm{s}_i \in \mathcal{S}_{\mathrm{test}}}
e^{i\Delta(\bm{s}_{i})}
\right|,
\]
provides a measure of predictive concentration. 
Values of $R_e$ close to $1$ indicate highly accurate predictions, 
whereas values near $0$ indicate dispersed or random errors.

\paragraph{Circular Correlation}

The agreement between predicted and observed angles may also be quantified using the Jammalamadaka--Sarma circular correlation:
\[
\rho_c
=
\frac{
\sum_{\bm{s}_i \in \mathcal{S}_{\mathrm{test}}}
\sin(Y(\bm{s}_i) - \bar{Y}_{\mathrm{test}})
\sin(\widehat{Y}(\bm{s}_i) - \bar{\widehat{Y}}_{\mathrm{test}})
}{
\sqrt{
\sum_{\bm{s}_i \in \mathcal{S}_{\mathrm{test}}}
\sin^2(Y(\bm{s}_i) - \bar{Y}_{\mathrm{test}})
\;
\sum_{\bm{s}_i \in \mathcal{S}_{\mathrm{test}}}
\sin^2(\widehat{Y}(\bm{s}_i) - \bar{\widehat{Y}}_{\mathrm{test}})
}
},
\]
where $\bar{Y}_{\mathrm{test}}$ and $\bar{\widehat{Y}}_{\mathrm{test}}$ denote circular means of $\{Y(\bm{s}_i), \bm{s}_i \in \mathcal{S}_{\mathrm{test}}\}$ and $\{\widehat{Y}(\bm{s}_i), \bm{s}_i \in \mathcal{S}_{\mathrm{test}}\}$. A $\rho_c \approx 1$ is preferable.

\paragraph{Average posterior concentration}

Let $\hat{c}(\bm s_i)$ denote the posterior predictive concentration (mean resultant length) at test location $\bm s_i$, for $\bm{s}_i \in \mathcal{S}_{\mathrm{test}}$. To summarize overall directional certainty across the test set, we compute the average posterior predictive concentration as
\begin{equation}
\nonumber \overline{c} = \frac{1}{\vert \mathcal{S}_{\mathrm{test}} \vert}
\sum_{\bm{s}_i \in \mathcal{S}_{\mathrm{test}}}
\hat{c}(\bm s_i).
\end{equation}
While this measure is comparable to $R_e$, it is based solely on the uncertainty of the posterior predictive distribution. A narrower credible interval is preferred in linear distributions, and similarly, a method providing a smaller value of $\overline{c}$ is preferred.

\subsection{Simulation setting and results}

We first fit the proposed model to the Indian Ocean wave direction dataset and obtain the posterior means. Accordingly, we choose the true parameter values in our simulation setting; we select them as $\mu_{\mathrm{true}} = 3$, $\sigma^2_{\mathrm{true}} = 10/3$, $\psi_{\mathrm{true}} = 3$, and $r_{\mathrm{true}}=0.95$ (not necessarily same as posterior means, but they are similar). Further, we use the same SPDE mesh shown in Figure \ref{fig:mesh_nodes} to draw 100 replications of dimension $N=33,845$ from the underlying GMRF in the process layer in \eqref{eq:final_model} and then wrap them to obtain 100 replications from the WGMRF model. As a result, all replications have a spatial correlation according to \eqref{cov_structure} and \eqref{eq:cicular_cor_pairwise}.

For validation, we split each replication into training and test sets at 80\% and 20\%, respectively. We use the \texttt{spatialBlock} function from the \texttt{blockCV} package in \texttt{R} to construct spatially structured cross-validation folds \citep{valavi2019blockcv}. The purpose of this function is to reduce bias arising from spatially correlated observations when they are randomly partitioned into training and test sets. Standard random cross-validation can lead to overly optimistic predictive performance because nearby observations, often highly correlated, may be split between training and validation subsets. The \texttt{spatialBlock} function partitions the study region into non-overlapping spatial blocks of a specified size or number. These blocks are typically constructed using square, rectangular, or user-defined polygons. Each block is then assigned to a fold, ensuring that all observations within the same spatial block belong to the same training or validation subset; this enforces spatial separation between training and testing data. Finally, after dividing the entire domain into five folds, we keep one fold for the testing set and the rest as the training set; finally, we have $\vert \mathcal{S}_{\mathrm{train}} \vert = 27,076$ and $\vert \mathcal{S}_{\mathrm{train}} \vert = 6,769$.

\begin{figure}[h]
    \centering
    \includegraphics[width=\textwidth]{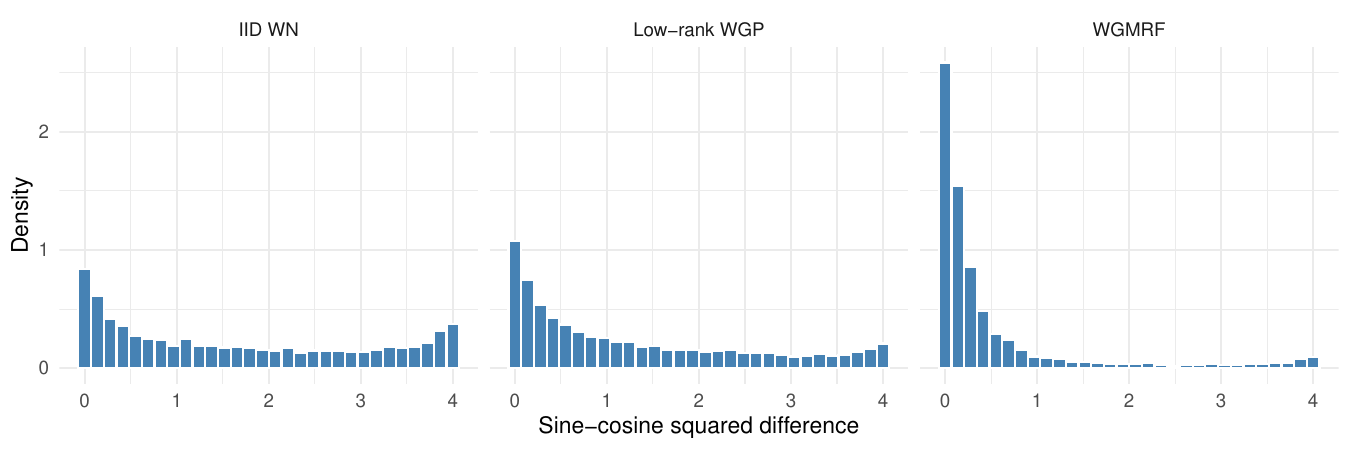}
    \includegraphics[width=\textwidth]{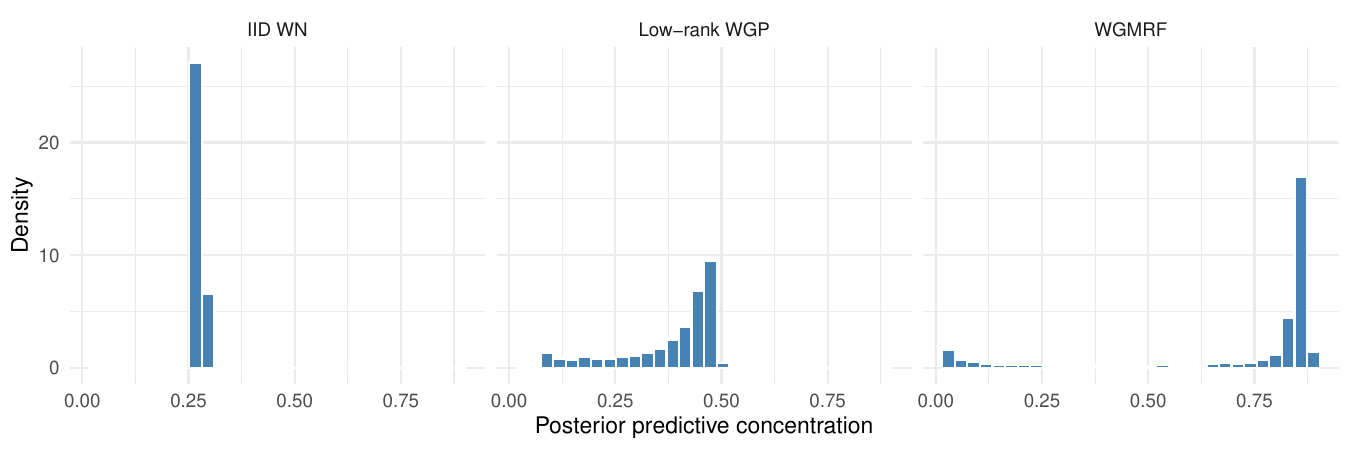}
    \caption{Top panel: Histograms of the pointwise sine-cosine squared differences $(\cos[Y(\bm{s}_i)] - \cos[\widehat{Y}(\bm{s}_i)])^2 + (\sin[Y(\bm{s}_i)] - \sin[\widehat{Y}(\bm{s}_i)])^2$ based on the non-spatial wrapped normal distribution (IID WN), the low-rank WGP in Section \ref{subsec:lowrank_wgp}, and the final proposed model (WGMRF) in \eqref{eq:final_model}. Bottom panel: Histograms of the pointwise posterior predictive concentration based on the same three models.}
    \label{fig:hist_square_diff_conc}
\end{figure}

We compare the proposed WGMRF model in \eqref{eq:final_model} with another proposed low-rank WGP model in Section \ref{subsec:lowrank_wgp} and the non-spatial wrapped normal distribution described in Section \ref{subsec:wn_univ}. For a representative replication, we show the histograms of the pointwise sine-cosine squared differences $(\cos[Y(\bm{s}_i)] - \cos[\widehat{Y}(\bm{s}_i)])^2 + (\sin[Y(\bm{s}_i)] - \sin[\widehat{Y}(\bm{s}_i)])^2$ in the top panel of Figure \ref{fig:hist_square_diff_conc}. We observe that the squared differences are most significant for the IID WN model, which does not account for spatial dependence. In the low-rank WGP model, the squared differences are lower than in the IID WN model. Further, for the proposed model, most squared differences are near zero, and the mass beyond the squared difference of 1 is negligible. We explore the results across all 100 replications and observe a similar pattern, suggesting consistent improvements in prediction performance for the proposed model. 

While the sine-cosine squared differences suggest lesser bias in prediction of the proposed approach, we further explore the distribution of the associated posterior concentration $\widehat{c}(\bm{s}_0)$ in \eqref{eq:krig_summary} for all three models based on the same replication as before; A value of posterior concentration near one ensures less prediction uncertainty. We show the histograms of the pointwise posterior concentration for all three competing models in the bottom panel of Figure \ref{fig:hist_square_diff_conc}. While the $\widehat{c}(\bm{s}_0)$ values are concentrated near 0.25 for the IID WN model, indicating high uncertainty in prediction, the values are mainly distributed between 0.1 and 0.5 for the low-rank WGP model. However, for the proposed model, most values are near 0.9, indicating a significantly better prediction performance of the WGMRF model in terms of lower prediction uncertainty. We again examine the results across all 100 replications and observe a similar pattern, suggesting a consistent lower prediction uncertainty for the proposed model.

We further explore the combined summary of the validation metrics described in Section \ref{subsec:validation_metrics}. These metrics provide a single-number summary of the model performance for each replication. We thus obtain 100 values for each metric; we further report their averages and standard errors in Table \ref{tab:comparison_results_sim}. Based on all metrics, the proposed WGMRF model outperforms the alternatives.

\begin{table}[ht]
\centering
\caption{Predictive performance comparison across IID WN, Low-rank WGP, and WGMRF models. 
Reported values are averages with standard errors shown in parentheses as subscripts (based on 100 replications of simulated datasets).}
\label{tab:comparison_results_sim}
\begin{tabular}{lccc}
\hline
 & IID WN & Low-rank WGP & WGMRF \\
\hline
SC-RMSE 
& $1.2693_{(0.0024)}$ 
& $1.1339_{(0.0029)}$ 
& $\mathbf{0.7314}_{(0.0033)}$ \\

$\bar{c}$ 
& $0.2848_{(0.0017)}$ 
& $0.3708_{(0.0020)}$ 
& $\mathbf{0.7106}_{(0.0023)}$ \\

CRMSE
& $1.5849_{(0.0038)}$ 
& $1.3768_{(0.0045)}$ 
& $\mathbf{0.8812}_{(0.0047)}$ \\

CMAE 
& $1.3237_{(0.0039)}$ 
& $1.1163_{(0.0042)}$ 
& $\mathbf{0.5931}_{(0.0033)}$ \\

$R_e$ 
& $0.1963_{(0.0031)}$ 
& $0.3580_{(0.0032)}$ 
& $\mathbf{0.7321}_{(0.0024)}$ \\

$\rho_c$ 
& $0.0019_{(0.0012)}$ 
& $0.4432_{(0.0034)}$ 
& $\mathbf{0.8364}_{(0.0021)}$ \\
\hline
\end{tabular}
\end{table}

\section{Data application}
\label{sec:application}

In this section, we first compare the proposed WGMRF with the alternatives using a 10-fold cross-validation. Further, using the best-performing models across all metrics in Section \ref{subsec:validation_metrics}, we draw statistical inferences.

\subsection{Model comparison}
\label{subsec:model_comparison}

Similar to simulation studies, we use the \texttt{spatialBlock} function from the \texttt{blockCV} package in \texttt{R} to construct the 10 folds of the observation locations. Subsequently, in a 10-fold cross-validation, we select one fold as the test set and the remaining nine as the training set, and finally repeat the study, treating each of the 10 folds as the test set. Here, under each fold, $\vert \mathcal{S}_{\mathrm{test}} \vert$ varies between 3384 and 3385.

\begin{figure}[t]
    \centering
    \includegraphics[width=\textwidth]{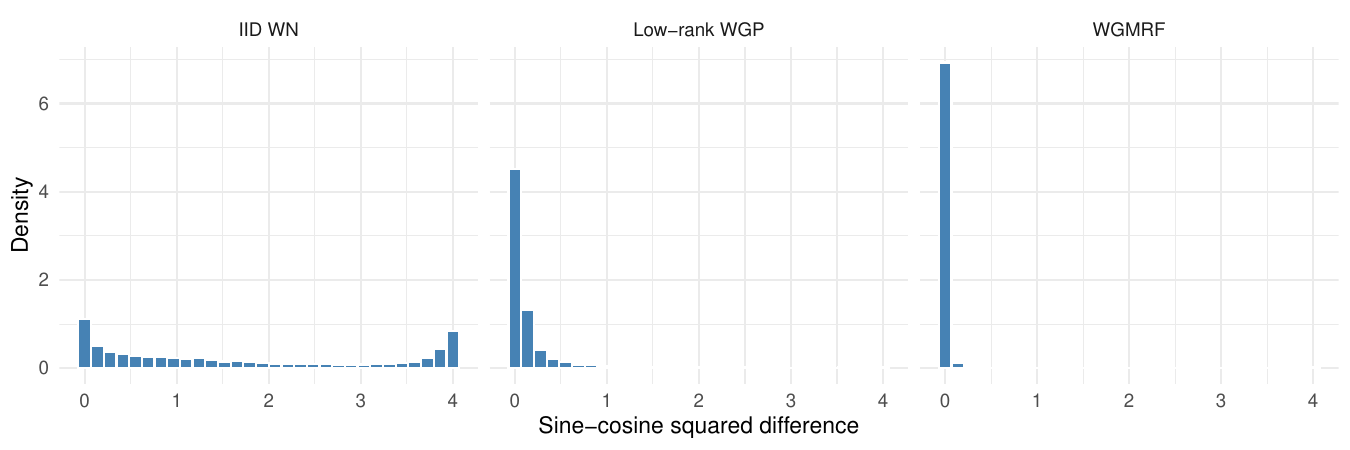}
    \includegraphics[width=\textwidth]{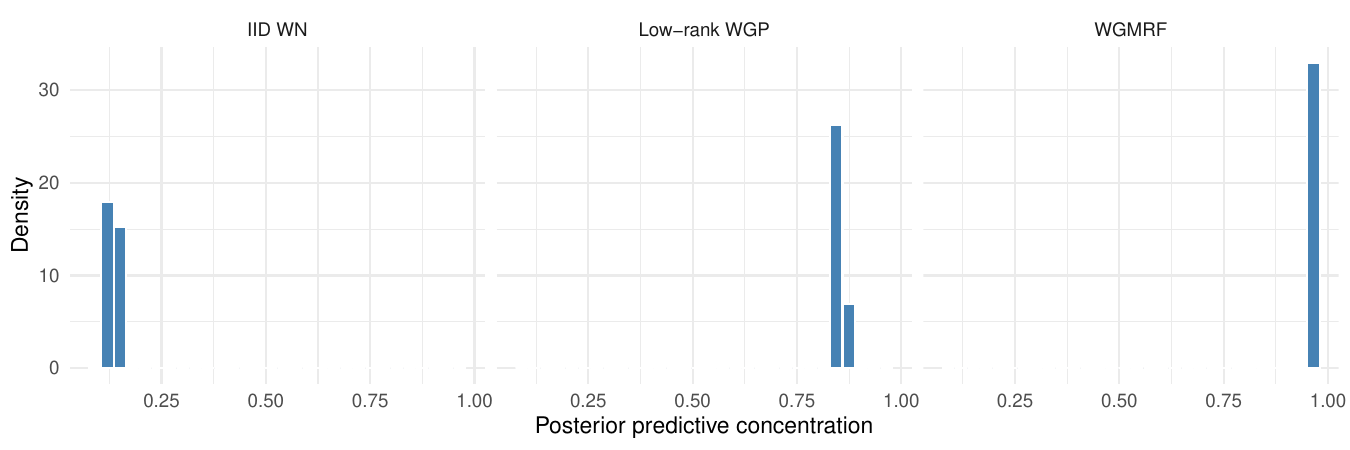}
    \caption{Top panel: Under a 10-fold cross-validation, histograms of the pointwise sine-cosine squared differences $(\cos[Y(\bm{s}_i)] - \cos[\widehat{Y}(\bm{s}_i)])^2 + (\sin[Y(\bm{s}_i)] - \sin[\widehat{Y}(\bm{s}_i)])^2$ based on the non-spatial wrapped normal distribution (IID WN), the low-rank WGP in Section \ref{subsec:lowrank_wgp}, and the final proposed model (WGMRF) in \eqref{eq:final_model}. Bottom panel: Histograms of the pointwise posterior predictive concentration under the same 10-fold cross-validation based on the same three models.}
    \label{fig:hist_square_diff_conc_crossval}
\end{figure}

Similar to the simulation studies, we compare the proposed WGMRF model in \eqref{eq:final_model} with the low-rank WGP model in Section \ref{subsec:lowrank_wgp} and the non-spatial wrapped normal distribution described in Section \ref{subsec:wn_univ}. The average (across folds) computation times for the non-spatial wrapped normal distribution, the low-rank WGP model, and the WGMRF model are 79.08 minutes, 172.65 minutes, and 133.03 minutes, respectively, i.e., the sparse precision matrix-based WGMRF model is even faster than the low-rank GMRF model with 100 knot locations. Combining all 10 folds, we show the histograms of the pointwise sine-cosine squared differences $(\cos[Y(\bm{s}_i)] - \cos[\widehat{Y}(\bm{s}_i)])^2 + (\sin[Y(\bm{s}_i)] - \sin[\widehat{Y}(\bm{s}_i)])^2$ in the top panel of Figure \ref{fig:hist_square_diff_conc_crossval}. We again observe that the squared differences are most significant for the IID WN model, which does not account for spatial dependence. Considering the low-rank WGP model, the squared differences are less than the IID WN model: while there is a spike near zero, it also exhibits a decaying tail between 0 and 1 on the $X$-axis. Further, in the proposed model, most of the squared differences are concentrated near zero, and the mass at higher squared differences is negligible. 

We further explore the distribution of the associated posterior concentration $\widehat{c}(\bm{s}_0)$ in \eqref{eq:krig_summary} for all three models based on the same 10-fold cross-validation. We show the histograms of the pointwise posterior concentration for all three competing models in the bottom panel of Figure \ref{fig:hist_square_diff_conc_crossval}. While the $\widehat{c}(\bm{s}_0)$ values are concentrated near 0.125 for the IID WN model, indicating high uncertainty in prediction, the values are concentrated near 0.85 for the low-rank WGP model, indicating a significant improvement over the IID WN model. However, for the proposed model, most values are near 0.97, suggesting even better prediction performance and lower prediction uncertainty.

We also explore the combined summary of the validation metrics described in Section \ref{subsec:validation_metrics}. These metrics provide a single-number summary of the model performance for each comparable-sized fold. We thus obtain 10 values for each metric; we further report their averages and standard errors in Table \ref{tab:comparison_results_crossval}. Based on all metrics, the proposed WGMRF model outperforms the alternatives.

\begin{table}[h]
\centering
\caption{Predictive performance comparison across IID, Low-rank WGP, and WGMRF models. 
Reported values are averages with standard errors shown in parentheses as subscripts  (based on 10 folds of a cross-validation of the wind direction dataset).}
\label{tab:comparison_results_crossval}
\begin{tabular}{lccc}
\hline
 & IID & Low-rank & SPDE \\
\hline
SC-RMSE
& $1.2979_{(0.0013)}$ 
& $0.4834_{(0.0010)}$ 
& $\mathbf{0.2255}_{(0.0009)}$ \\

$\bar{c}$
& $0.1350_{(0.0001)}$ 
& $0.8558_{(0.0001)}$ 
& $\mathbf{0.9667}_{(0.0003)}$ \\

CRMSE
& $1.7004_{(0.0019)}$ 
& $0.5591_{(0.0014)}$ 
& $\mathbf{0.2885}_{(0.0013)}$ \\

CMAE 
& $1.3805_{(0.0021)}$ 
& $0.3314_{(0.0006)}$ 
& $\mathbf{0.0674}_{(0.0003)}$ \\

$R_e$
& $0.1579_{(0.0016)}$ 
& $0.8832_{(0.0005)}$ 
& $\mathbf{0.9746}_{(0.0002)}$ \\

$\rho_c$
& $0.0052_{(0.0012)}$ 
& $0.7843_{(0.0023)}$ 
& $\mathbf{0.9615}_{(0.0008)}$ \\
\hline
\end{tabular}
\end{table}

\subsection{Results based on WGMRF model}


Given the superior performance of the proposed WGMRF model over its alternatives, we draw inferences for the wind direction dataset over the entire Indian Ocean basin for the hour of the 2004 Indian Ocean Tsunami. As already mentioned in Section \ref{subsec:bayescomp}, we obtain a total of 60,000 MCMC samples, and discard the first 10,000 samples in the burn-in phase, and after thinning by a length of five, we finally draw posterior inferences based on 10,000 remaining samples. The computation time on an AMD Ryzen 9 5900X processor with 64GB RAM is 316.16 minutes, which is highly moderate considering the $N=33,845$ data locations. 

\begin{figure}[h]
    \centering
    \includegraphics[width=\textwidth]{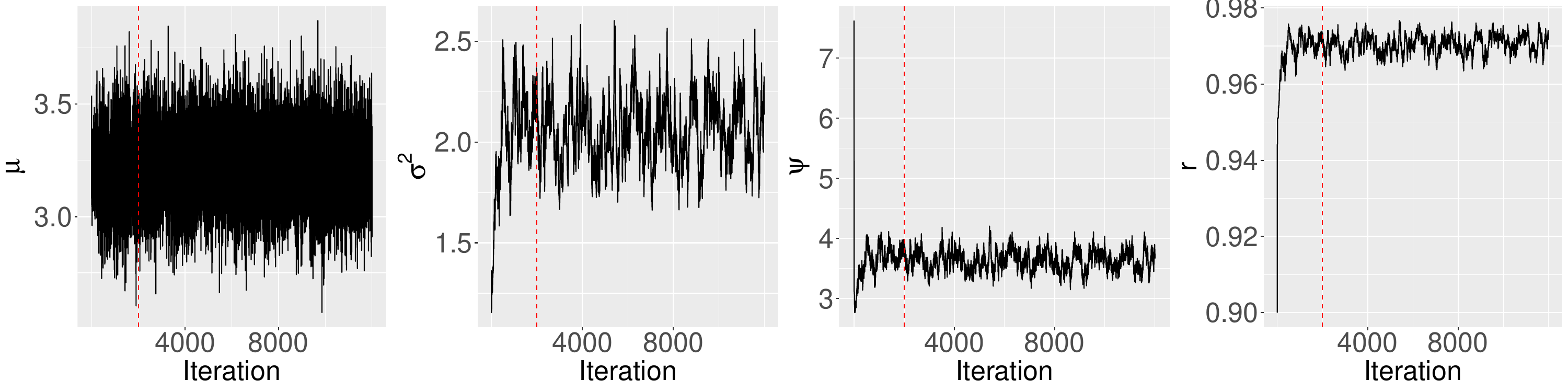}
    \includegraphics[width=\textwidth]{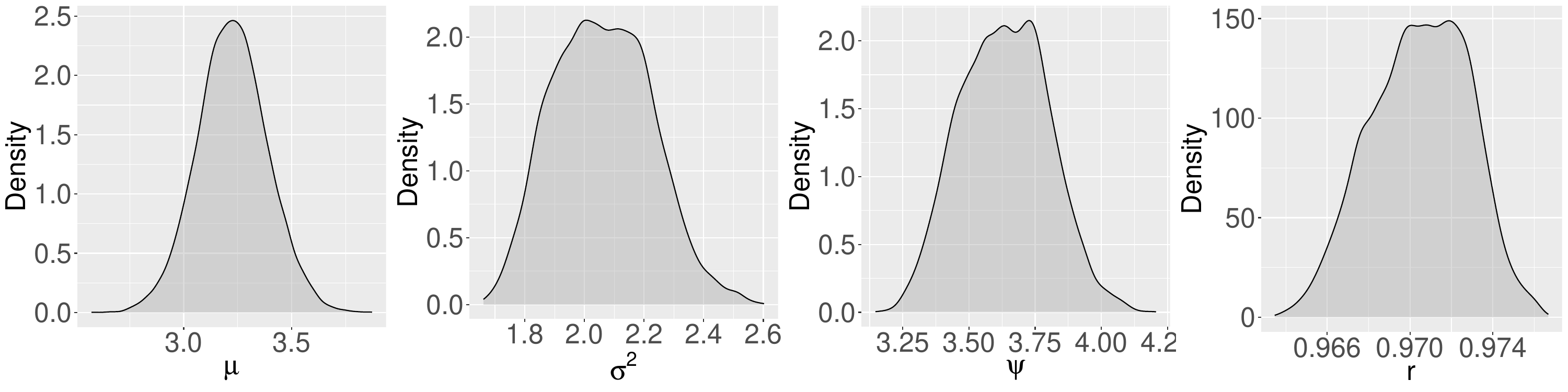}
    \caption{Top panel: Trace plots of the model parameters $\mu$, $\sigma^2$, $\psi$, and $\nu$. We show the MCMC chains of length 12,000, including the burn-in and post-burn-in phases, but after thinning by five. Bottom panel: Posterior densities of the parameters, approximated based on the post-burn-in samples.}
    \label{fig:traceplots_densities}
\end{figure}

The trace plots of the model parameters and the kernel density estimates of their posterior distributions based on post-burn-in samples are provided in Figure \ref{fig:traceplots_densities}. The convergence appears reasonable for all chains; we start from values far from the bulk of the posterior distributions for $\sigma^2$, $\psi$, and $r$, and they approach near the correct stationary distribution fast. The posterior densities are bell-shaped and appear significantly different from their non-informative prior distributions, indicating Bayesian learning.

\begin{figure}[h]
    \centering
    \includegraphics[height=0.35\linewidth]{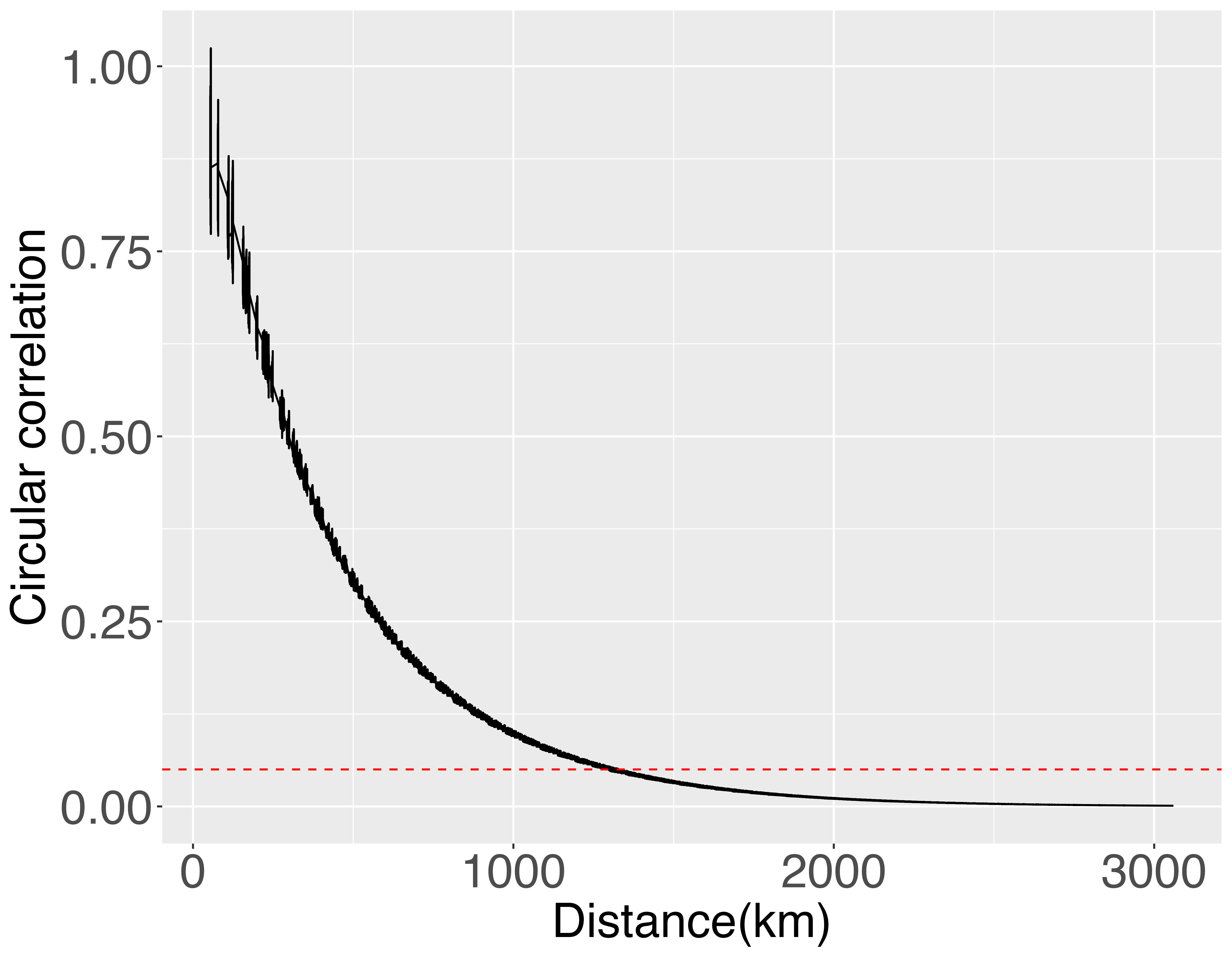}
    \includegraphics[height=0.35\textwidth]{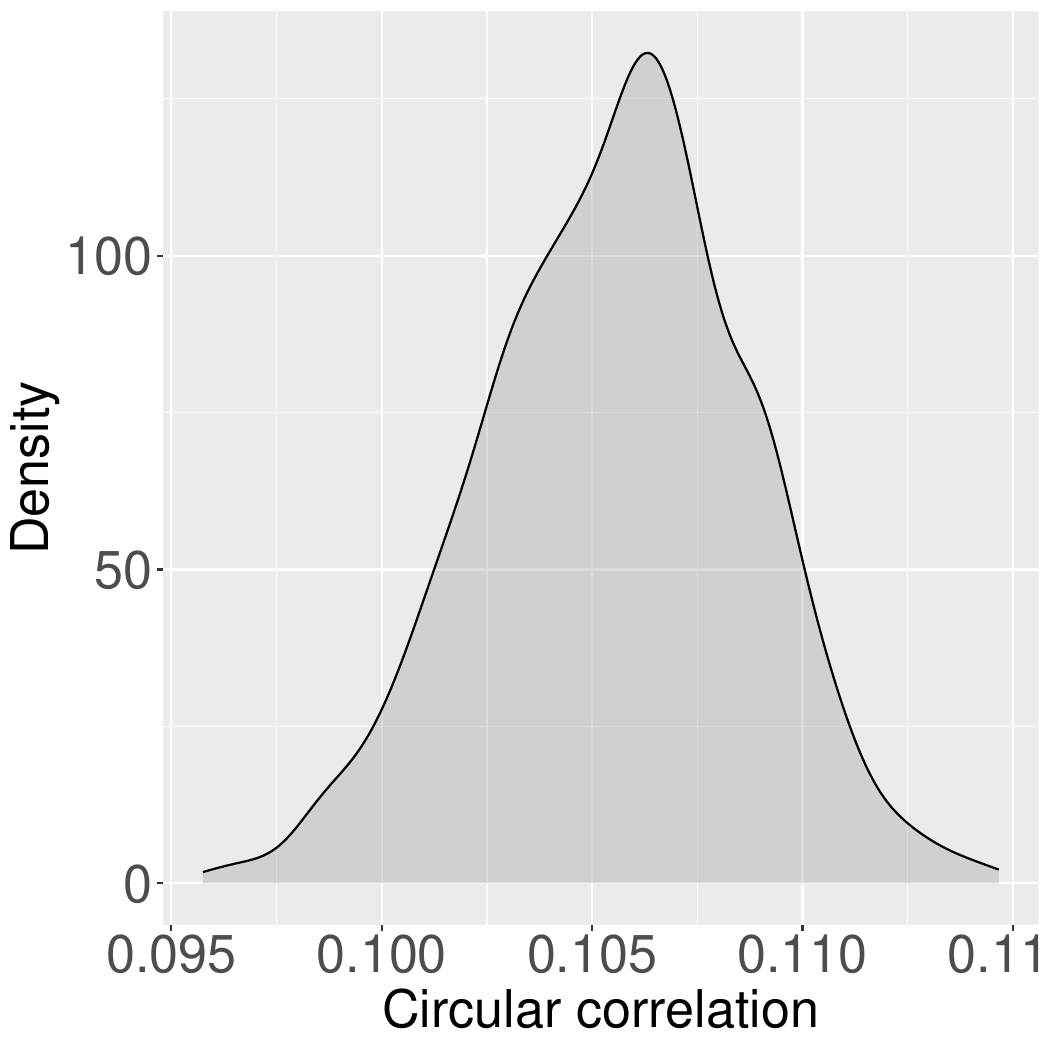}
    \caption{Left panel: Geodesic distance (in kilometers) versus the estimated spatial circular correlation using \eqref{eq:cicular_cor_pairwise} and \eqref{eq:approx_matern_cormat}. The horizontal dashed line represents a circular correlation of 0.05; the corresponding value on the X-axis indicates the spatial circular range. Right panel: The posterior distribution of the spatial circular correlation between two grid cells at ($92.5^\circ$E, $11.5^\circ$N) and ($96.0^\circ$E, $3.5^\circ$N).}
    \label{fig:est_spatial_circcor}
\end{figure}

Further, we explore the fitted spatial correlation of the wind direction data. Here, for each set of MCMC sample $\{\mu^{(b)}, \sigma^{2(b)}, \psi^{(b)}, r^{(b)} \}$, where $b=1,\ldots,B$, we need to evaluate the $(N\times N)$-dimensional matrix $\tilde{\bm{\Sigma}}_{\psi, r}$ in \eqref{eq:approx_matern_cormat} and further need to calculate the inherited circular correlation in \eqref{eq:cicular_cor_pairwise} for each pair of elements in $\tilde{\bm{\Sigma}}_{\psi, r}$. Further, averaging them across the $B=10,000$ remaining post-burn-in samples yields the final SPDE-approximated circular spatial correlation, which is challenging from both computational and storage perspectives. Given that our main aim is to understand how the spatial circular correlation decays with distance, we first consider a $10^\circ \times 10^\circ$ window centered on the tsunami epicenter at $95.854^\circ$E and $3.316^\circ$N. Within this region, there are 1,356 grid cells. We evaluate the SPDE-approximated circular spatial correlation only for each pair of those locations. Further, we compare the geodesic distances between the locations and the corresponding circular correlation values in the left panel of Figure \ref{fig:est_spatial_circcor}. A few estimated correlation values exceed 1 due to the SPDE-based approximation. The effective spatial range is usually defined as the distance at which the spatial correlation decays to 0.05. Similarly, we define the distance at which the spatial circular correlation decays to 0.05 as the effective spatial circular range. Here, the estimated spatial circular range is approximately 1327.72 kilometers. Overall, we observe a strong circular correlation across a high range of distances. Given that our methodology assumes the underlying spatial correlation to be isotropic, the estimates would be (approximately, due to the SPDE-based approximation) equal between any pairs of grid cells at similar geodesic distances.

While the above-mentioned approach provides point estimates of spatial circular correlation, i.e., the element-wise posterior means, we can obtain the full posterior distribution of the spatial circular correlation for a pair of locations as well. For demonstration, we select two grid cells, one with the centroid nearest to the epicenter, i.e., $96.0^\circ$E, $3.5^\circ$N. The other one is with the centroid at $92.5^\circ$E and $11.5^\circ$N, nearest to Port Blair, the capital of the Andaman and Nicobar Islands, India, with actual coordinates $92.7265^\circ$E and $11.6234^\circ$N. The geodesic distance between the centroids of these two grid cells is 970.63 kilometers. The posterior distribution of the spatial circular correlation between these two grid cells is presented in the right panel of Figure \ref{fig:est_spatial_circcor}. The distribution is bell-shaped, with a Monte Carlo posterior mean of 0.1056 and a posterior standard deviation of 0.0031.

\section{Conclusion}
\label{sec:conclusion}

In this paper, we propose a wrapped Gaussian Markov random field model for analyzing large spatially correlated circular datasets, with a focus on an ERA reanalysis-based wind direction dataset spanning the entire Indian Ocean basin corresponding to the hour of the 2004 Indian Ocean Tsunami. The proposed model exploits the idea of wrapping a dense Gaussian process model proposed by \cite{jona2012spatial} and the explicit link between a dense Gaussian process and a Gaussian Markov random field with a sparse precision matrix proposed by \cite{lindgren2011explicit}. Here, we approximate a dense Mat\'ern correlation structure using a stochastic partial differential equation-based construction for the latent Gaussian process. In our model, it is not necessary to explicitly define a correlation structure for the data layer. Instead, the data layer inherits the spatial dependence structure of the latent Gaussian Markov random field that allows sparse precision matrices. The computation for the proposed model is feasible using Markov chain Monte Carlo sampling, with latent winding numbers updated quickly via parallel updates, and the latent spatial random effects can also be updated quickly due to the sparse precision matrix. Through an extensive simulation study and an application to the 2004 Indian Ocean Tsunami dataset, we demonstrate that the proposed model provides improved predictive performance relative to competing approaches. The exploratory analysis of the wind direction data during the tsunami reveals strong spatial correlation, and the proposed wrapped GMRF framework effectively captures this structure while remaining scalable across large spatial domains.

Several avenues for future research arise naturally from this work. A primary direction is the development of spatiotemporal wrapped Gaussian process models suitable for large-scale dynamic circular datasets. Extending the SPDE-based construction to spatiotemporal settings would allow coherent modeling of evolving directional fields while maintaining computational scalability. More broadly, numerous approximation strategies exist for large spatial Gaussian process models, including Vecchia-type approximations and nearest-neighbor Gaussian processes (NNGP); a detailed discussion and comparison of these approaches is in \cite{hazra2025exploring}. It would be of considerable interest to investigate how such approximation schemes can be adapted to the wrapped setting, thereby yielding additional classes of scalable models for massive spatial circular datasets. Exploring theoretical properties, computational trade-offs, and predictive performance of these wrapped approximations represents a promising direction for future research. Finally, further methodological developments may focus on incorporating covariates, modeling anisotropy on the sphere, and related areas. Such extensions would enhance the applicability of wrapped spatial models in geophysical, environmental, and climate science applications where large-scale circular data arise naturally.


\section*{Data availability}
The data and code underlying this article are provided at the GitHub link \url{https://github.com/arnabstatswithR/WGMRF.git}. 

\bibliographystyle{plainnat} 
\bibliography{reference}          

@article{jona2012spatial,
  title={{Spatial analysis of wave direction data using wrapped Gaussian processes}},
  author={Jona-Lasinio, Giovanna and Gelfand, Alan and Jona-Lasinio, Mattia},
  volume={6},
  number={4},
  pages={1478--1498},
  year={2012}
}

@article{pewsey2021recent,
  title={Recent advances in directional statistics},
  author={Pewsey, Arthur and Garc{\'\i}a-Portugu{\'e}s, Eduardo},
  journal={Test},
  volume={30},
  number={1},
  pages={1--58},
  year={2021},
  publisher={Springer}
}

@inproceedings{rezende2020normalizing,
  title={Normalizing flows on tori and spheres},
  author={Rezende, Danilo Jimenez and Papamakarios, George and Racaniere, S{\'e}bastien and Albergo, Michael and Kanwar, Gurtej and Shanahan, Phiala and Cranmer, Kyle},
  booktitle={International Conference on Machine Learning},
  pages={8083--8092},
  year={2020},
  organization={PMLR}
}

@article{gelfand2016spatial,
  title={{Spatial statistics and Gaussian processes: A beautiful marriage}},
  author={Gelfand, Alan E and Schliep, Erin M},
  journal={Spatial Statistics},
  volume={18},
  pages={86--104},
  year={2016},
  publisher={Elsevier}
}

@article{arrieta2024spatially,
  title={Spatially transferable machine learning wind power prediction models: v-logit random forests},
  author={Arrieta-Prieto, Mario and Schell, Kristen R},
  journal={Renewable Energy},
  volume={223},
  pages={120066},
  year={2024},
  publisher={Elsevier}
}

@article{wang2014modeling,
  title={{Modeling space and space-time directional data using projected Gaussian processes}},
  author={Wang, Fangpo and Gelfand, Alan E},
  journal={Journal of the American Statistical Association},
  volume={109},
  number={508},
  pages={1565--1580},
  year={2014},
  publisher={Taylor \& Francis}
}

@inproceedings{mallasto2018wrapped,
  title={{Wrapped Gaussian process regression on Riemannian manifolds}},
  author={Mallasto, Anton and Feragen, Aasa},
  booktitle={Proceedings of the IEEE Conference on Computer Vision and Pattern Recognition},
  pages={5580--5588},
  year={2018}
}

@article{mastrantonio2016spatio,
  title={Spatio-temporal circular models with non-separable covariance structure},
  author={Mastrantonio, Gianluca and Jona Lasinio, Giovanna and Gelfand, Alan E},
  journal={Test},
  volume={25},
  number={2},
  pages={331--350},
  year={2016},
  publisher={Springer}
}

@article{wang2015joint,
  title={{Joint spatio-temporal analysis of a linear and a directional variable: space-time modeling of wave heights and wave directions in the Adriatic Sea}},
  author={Wang, Fangpo and Gelfand, Alan E and Jona-Lasinio, Giovanna},
  journal={Statistica Sinica},
  pages={25--39},
  year={2015},
  publisher={JSTOR}
}

@article{marques2022non,
  title={{A non-stationary model for spatially dependent circular response data based on wrapped Gaussian processes}},
  author={Marques, Isa and Kneib, Thomas and Klein, Nadja},
  journal={Statistics and Computing},
  volume={32},
  number={5},
  pages={73},
  year={2022},
  publisher={Springer}
}

@article{lindgren2011explicit,
  title={{An explicit link between Gaussian fields and Gaussian Markov random fields: the stochastic partial differential equation approach}},
  author={Lindgren, Finn and Rue, H{\aa}vard and Lindstr{\"o}m, Johan},
  journal={Journal of the Royal Statistical Society Series B: Statistical Methodology},
  volume={73},
  number={4},
  pages={423--498},
  year={2011},
  publisher={Oxford University Press}
}

@article{hazra2021estimating,
  title={{Estimating high-resolution Red Sea surface temperature hotspots, using a low-rank semiparametric spatial model}},
  author={Hazra, Arnab and Huser, Rapha{\"e}l},
  journal={The Annals of Applied Statistics},
  volume={15},
  number={2},
  pages={572--596},
  year={2021},
  publisher={Institute of Mathematical Statistics}
}

@article{hazra2025efficient,
  title={Efficient modeling of spatial extremes over large geographical domains},
  author={Hazra, Arnab and Huser, Rapha{\"e}l and Bolin, David},
  journal={Journal of Computational and Graphical Statistics},
  volume={34},
  number={3},
  pages={795--811},
  year={2025},
  publisher={Taylor \& Francis}
}

@incollection{hazra2023bayesian,
  title={{Bayesian latent Gaussian models for high-dimensional spatial extremes}},
  author={Hazra, Arnab and Huser, Rapha{\"e}l and J{\'o}hannesson, {\'A}rni V},
  booktitle={Statistical Modeling Using Bayesian Latent Gaussian Models: With Applications in Geophysics and Environmental Sciences},
  pages={219--251},
  year={2023},
  publisher={Springer}
}

@article{hersbach2020era5,
  title={{The ERA5 global reanalysis}},
  author={Hersbach, Hans and Bell, Bill and Berrisford, Paul and Hirahara, Shoji and Hor{\'a}nyi, Andr{\'a}s and Mu{\~n}oz-Sabater, Joaqu{\'\i}n and Nicolas, Julien and Peubey, Carole and Radu, Raluca and Schepers, Dinand and others},
  journal={Quarterly Journal of the Royal Meteorological Society},
  volume={146},
  number={730},
  pages={1999--2049},
  year={2020},
  publisher={Wiley Online Library}
}

@book{holthuijsen2010waves,
  title={Waves in oceanic and coastal waters},
  author={Holthuijsen, Leo H},
  year={2010},
  publisher={Cambridge University Press}
}

@article{young2011global,
  title={Global trends in wind speed and wave height},
  author={Young, IR and Zieger, Stefan and Babanin, Alexander V},
  journal={Science},
  volume={332},
  number={6028},
  pages={451--455},
  year={2011},
  publisher={American Association for the Advancement of Science}
}

@article{young2019multiplatform,
  title={Multiplatform evaluation of global trends in wind speed and wave height},
  author={Young, Ian R and Ribal, Agustinus},
  journal={Science},
  volume={364},
  number={6440},
  pages={548--552},
  year={2019},
  publisher={American Association for the Advancement of Science}
}

@article{lay2005great,
  title={{The great Sumatra-Andaman earthquake of 26 December 2004}},
  author={Lay, Thorne and Kanamori, Hiroo and Ammon, Charles J and Nettles, Meredith and Ward, Steven N and Aster, Richard C and Beck, Susan L and Bilek, Susan L and Brudzinski, Michael R and Butler, Rhett and others},
  journal={Science},
  volume={308},
  number={5725},
  pages={1127--1133},
  year={2005},
  publisher={American Association for the Advancement of Science}
}

@article{ashton2001formation,
  title={Formation of coastline features by large-scale instabilities induced by high-angle waves},
  author={Ashton, Andrew and Murray, A Brad and Arnoult, Olivier},
  journal={Nature},
  volume={414},
  number={6861},
  pages={296--300},
  year={2001},
  publisher={Nature Publishing Group UK London}
}

@article{watson1961goodness,
 author = {G. S. Watson},
 journal = {Biometrika},
 number = {1/2},
 pages = {109--114},
 publisher = {[Oxford University Press, Biometrika Trust]},
 title = {Goodness-Of-Fit Tests on a Circle},
 urldate = {2026-02-24},
 volume = {48},
 year = {1961}
}

@article{watson1962goodness,
 author = {G. S. Watson},
 journal = {Biometrika},
 number = {1/2},
 pages = {57--63},
 publisher = {[Oxford University Press, Biometrika Trust]},
 title = {Goodness-of-Fit Tests on a Circle. II},
 urldate = {2026-02-24},
 volume = {49},
 year = {1962}
}

@article{stephens1963random,
 author = {M. A. Stephens},
 journal = {Biometrika},
 number = {3/4},
 pages = {385--390},
 publisher = {[Oxford University Press, Biometrika Trust]},
 title = {Random Walk on a Circle},
 urldate = {2026-02-24},
 volume = {50},
 year = {1963}
}

@article{mardia2008multivariate,
  title={{A multivariate von Mises distribution with applications to bioinformatics}},
  author={Mardia, Kanti V and Hughes, Gareth and Taylor, Charles C and Singh, Harshinder},
  journal={Canadian Journal of Statistics},
  volume={36},
  number={1},
  pages={99--109},
  year={2008},
  publisher={Wiley Online Library}
}

@article{stephens1970use,
  title={{Use of the Kolmogorov--Smirnov, Cramer--Von Mises and related statistics without extensive tables}},
  author={Stephens, Michael A},
  journal={Journal of the Royal Statistical Society Series B: Statistical Methodology},
  volume={32},
  number={1},
  pages={115--122},
  year={1970},
  publisher={Oxford University Press}
}

@article{mardia1975statistics,
  title={Statistics of directional data},
  author={Mardia, Kantilal Varichand},
  journal={Journal of the Royal Statistical Society Series B: Statistical Methodology},
  volume={37},
  number={3},
  pages={349--371},
  year={1975},
  publisher={Oxford University Press}
}

@book{mardia2000directional,
    AUTHOR = {Mardia, Kanti V. and Jupp, Peter E.},
     TITLE = {Directional statistics},
    SERIES = {Wiley Series in Probability and Statistics},
      NOTE = {Revised reprint of {\it Statistics of directional data} by
              Mardia [MR0336854 (49 \#1627)]},
 PUBLISHER = {John Wiley \& Sons, Ltd., Chichester},
      YEAR = {2000},
     PAGES = {xxii+429},
      ISBN = {0-471-95333-4},
   MRCLASS = {62-02 (62E99 62H11)},
  MRNUMBER = {1828667},
MRREVIEWER = {Richard\ A.\ Johnson},
}

@article{harrison1988development,
  title={The development of analysis of variance for circular data},
  author={Harrison, D and Kanji, GK},
  journal={Journal of Applied Statistics},
  volume={15},
  number={2},
  pages={197--223},
  year={1988},
  publisher={Taylor \& Francis}
}

@article{damien1999full,
  title={{A full Bayesian analysis of circular data using the von Mises distribution}},
  author={Damien, Paul and Walker, Stephen},
  journal={The Canadian Journal of Statistics/La Revue Canadienne de Statistique},
  pages={291--298},
  volume={27},
  year={1999},
  publisher={JSTOR}
}

@article{kato2008circular,
  title={A circular--circular regression model},
  author={Kato, Shogo and Shimizu, Kunio and Shieh, Grace S},
  journal={Statistica Sinica},
  pages={633--645},
  volume={18},
  number={2},
  year={2008},
  publisher={JSTOR}
}

@article{coles1998inference,
  title={Inference for circular distributions and processes},
  author={Coles, Stuart},
  journal={Statistics and Computing},
  volume={8},
  number={2},
  pages={105--113},
  year={1998},
  publisher={Springer}
}

@article{hazra2025exploring,
  title={Exploring the efficacy of statistical and deep learning methods for large spatial datasets: A case study},
  author={Hazra, Arnab and Nag, Pratik and Yadav, Rishikesh and Sun, Ying},
  journal={Journal of Agricultural, Biological and Environmental Statistics},
  volume={30},
  number={1},
  pages={231--254},
  year={2025},
  publisher={Springer}
}

@incollection{higdon2002space,
	Author = {Higdon, Dave},
	Booktitle = {Quantitative methods for current environmental issues},
	Pages = {37--56},
	Publisher = {Springer},
	Title = {Space and space-time modeling using process convolutions},
	Year = {2002}}

@article{wikle1999dimension,
	Author = {Wikle, Christopher K and Cressie, Noel},
	Journal = {Biometrika},
	Number = {4},
	Pages = {815--829},
	Publisher = {Oxford University Press},
	Title = {A dimension-reduced approach to space-time {K}alman filtering},
	Volume = {86},
	Year = {1999}}

@article{banerjee2008gaussian,
	Author = {Banerjee, Sudipto and Gelfand, Alan E and Finley, Andrew O and Sang, Huiyan},
	Journal = {Journal of the Royal Statistical Society: Series B (Statistical Methodology)},
	Number = {4},
	Pages = {825--848},
	Publisher = {Wiley Online Library},
	Title = {Gaussian predictive process models for large spatial data sets},
	Volume = {70},
	Year = {2008}}

@article{fuentes2007approximate,
	Author = {Fuentes, Montserrat},
	Journal = {Journal of the American Statistical Association},
	Number = {477},
	Pages = {321--331},
	Publisher = {Taylor \& Francis},
	Title = {Approximate likelihood for large irregularly spaced spatial data},
	Volume = {102},
	Year = {2007}}

@article{vecchia1988estimation,
	Author = {Vecchia, Aldo V},
	Journal = {{Journal of the Royal Statistical Society: Series B (Statistical Methodology)}},
	Number = {2},
	Pages = {297--312},
	Title = {Estimation and model identification for continuous spatial processes},
	Volume = {50},
	Year = {1988}}

@article{stein2004approximating,
	Author = {Stein, Michael L and Chi, Zhiyi and Welty, Leah J},
	Journal = {Journal of the Royal Statistical Society: Series B (Statistical Methodology)},
	Number = {2},
	Pages = {275--296},
	Publisher = {Wiley Online Library},
	Title = {Approximating likelihoods for large spatial data sets},
	Volume = {66},
	Year = {2004}}

@article{furrer2006covariance,
	Author = {Furrer, Reinhard and Genton, Marc G and Nychka, Douglas},
	Journal = {Journal of Computational and Graphical Statistics},
	Number = {3},
	Pages = {502--523},
	Publisher = {Taylor \& Francis},
	Title = {Covariance tapering for interpolation of large spatial datasets},
	Volume = {15},
	Year = {2006}}

@book{rue2005gaussian,
	Author = {Rue, H{\aa}vard and Held, Leonhard},
	Publisher = {Chapman and Hall/CRC},
	Title = {{Gaussian Markov Random Fields: Theory and Applications}},
	Year = {2005}}

@article{ishii2005extent,
  title={{Extent, duration and speed of the 2004 Sumatra--Andaman earthquake imaged by the Hi-Net array}},
  author={Ishii, Miaki and Shearer, Peter M and Houston, Heidi and Vidale, John E},
  journal={Nature},
  volume={435},
  number={7044},
  pages={933--936},
  year={2005},
  publisher={Nature Publishing Group UK London}
}

@article{cisneros2023combined,
  title={A combined statistical and machine learning approach for spatial prediction of extreme wildfire frequencies and sizes},
  author={Cisneros, Daniela and Gong, Yan and Yadav, Rishikesh and Hazra, Arnab and Huser, Rapha{\"e}l},
  journal={Extremes},
  volume={26},
  number={2},
  pages={301--330},
  year={2023},
  publisher={Springer}
}

@article{sahoo2025computationally,
  title={Computationally scalable Bayesian SPDE modeling for censored spatial responses},
  author={Sahoo, Indranil and Majumder, Suman and Hazra, Arnab and Rappold, Ana G and Bandyopadhyay, Dipankar},
  journal={The New England Journal of Statistics in Data Science},
  year={2025},
  pages={1--15},
  doi={10.51387/25-NEJSDS78}
}

@article{valavi2019blockcv,
  title={{blockCV: An R package for generating spatially or environmentally separated folds for k-fold cross-validation of species distribution models}},
  author={Valavi, Roozbeh and Elith, Jane and Lahoz-Monfort, Jos{\'e} J and Guillera-Arroita, Gurutzeta},
  journal = {Methods in Ecology and Evolution},
    volume = {10},
    number = {2},
    pages = {225-232},
  year={2019}
}

@inproceedings{jammalamadaka1988correlation,
  title={{A correlation coefficient for angular variables. Statistical Theory and Data Analysis 2}},
  author={Jammalamadaka, S and Sarma, Y},
  booktitle={Proceedings of the Second Pacific Area Statistical Conference. Amsterdam, The Netherlands: North Holland},
  pages={349--364},
  year={1988}
}

@book{jammalamadaka2001topics,
  title={Topics in circular statistics},
  author={Jammalamadaka, S Rao and Sengupta, Ambar},
  volume={5},
  year={2001},
  publisher={world scientific}
}

@book{fisher1995statistical,
  title={Statistical Analysis of Circular Data},
  author={Fisher, Nicholas I},
  year={1995},
  publisher={Cambridge University Press}
}

\end{document}